\documentclass[reprint,amsmath,amssymb,aps,pre,numeric,nofootinbib, graphics,floatfix]{revtex4-1}
\usepackage{subcaption}
\usepackage{tabularx}
\usepackage{graphicx}% Include figure files
\usepackage{color}% Inlcude figures with color
\usepackage{dcolumn}% Align table columns on decimal point
\usepackage{epstopdf}% Making epslatex possible
\usepackage{bm}% bold math
\usepackage[utf8]{inputenc}
\usepackage{mathtools}
\usepackage{xcolor}
\usepackage{xcolor}

\usepackage{natbib} % this is probably into you nips_2018 format file

\begin{document}

\preprint{APS/123-QED}

\title{Machine Learning as an Accurate  Predictor for Percolation Threshold of Diverse Networks}% Force line breaks with \\

\author{Siddharth Patwardhan\textsuperscript{$a,b$}, Utso Majumder\textsuperscript{$c$}, Aditya Das Sarma\textsuperscript{$c$},  
Mayukha Pal\textsuperscript{$d$},
Divyanshi Dwivedi\textsuperscript{$d,e$}, and
Prasanta K. Panigrahi\textsuperscript{$f,*$}}

\affiliation{\textsuperscript{$a$}Department of Mathematics and Statistics, Indian Institute of Science Education and Research, Kolkata,  741246, West Bengal, India}
\affiliation{\textsuperscript{$b$}Center for Complex Networks and Systems Research, IU-Bloomington, IN 47405, USA}
\affiliation{\textsuperscript{$c$}Department of Electronics \& Telecommunication Engineering, Jadavpur University, Kolkata - 700032, West Bengal, India}
\affiliation{\textsuperscript{$d$}ABB Ability Innovation Center,  Asea Brown Boveri Company, Hyderabad, 500084, Telangana, India}
\affiliation{\textsuperscript{$e$}Department of Electrical Engineering, Indian Institutes of Technology- Hyderabad, 502205, Telangana, India}
\affiliation{\textsuperscript{$f$}Department of Physical Sciences, Indian Institute of Science Education and Research, Kolkata, 741246, West Bengal, India}
\affiliation{\textsuperscript{$*$} Corresponding Author, Email Address: pprasanta@iiserkol.ac.in}

\begin{abstract}
The percolation threshold is an important measure to determine the inherent rigidity of large networks. Predictors of the percolation threshold for large networks are computationally intense to run, hence it is a necessity to develop predictors of the percolation threshold of networks, that do not rely on numerical simulations. We demonstrate the efficacy of five machine learning-based regression techniques for the accurate prediction of the percolation threshold. The dataset generated to train the machine learning models contains a total of 777 real and synthetic networks. It consists of 5 statistical and structural properties of networks as features and the numerically computed percolation threshold as the output attribute. We establish that the machine learning models outperform three existing empirical estimators of bond percolation threshold, and extend this experiment to predict site and explosive percolation. Further, we compared the performance of our models in predicting the percolation threshold using RMSE values. The gradient boosting regressor, multilayer perceptron and random forests regression models achieve the least RMSE values among considered models.
\end{abstract}

\maketitle

\section{Introduction}

Percolation theory is an important domain of statistical physics and mathematics, it describes the behaviour of a network on the addition or modification of nodes. Applications of percolation theory to material science and in many other disciplines have been extensively discussed. Percolation, in particular, is the most extensively evaluated process in statistical physics \cite{per}. 

There exists an extensive study on the responsive behaviour of complex networks when subjected to any strike on the vertices and edges. To investigate the available complex networks and real-world networks such as scientific collaboration and Internet traffic; a quantifiable measure based on average inverse geodesic distance and size of the largest connected clusters is used for analyzing the network performance. There are four ways to attack the vertices and edges of the network which are as follows: removal of vertices and edges in decreasing order of
\begin{itemize}
    \item Degree in the elementary network,
    \item Betweenness centrality in the elementary network, 
    \item Degree in the current network, and
    \item Betweenness centrality in the current network.
\end{itemize}

It has been observed that an attacking strategy based on the elementary network is not reliable as the network structure gets transformed because of the removal of well-connected vertices or edges. Furthermore, for complex networks, the correlation between the degree and betweenness centrality measures is analyzed.

The percolation threshold is denoted by $\rho_c$, referred to as the critical circumstance of the model. In a model, if the probability is less than the percolation threshold i.e., $p < \rho_c$, then it is referred to as a subcritical percolation. On the other hand, if $p > \rho_c$ it is referred to as a supercritical percolation. Because of this distinction, the percolation threshold ($\rho_c$) is represented as the phase transition of the model since it is a point exactly in between subcritical and supercritical phases. It is to be noted that subcritical percolation models should not have infinite connected components, while supercritical models always consist of at least one such component.

However, the computation of the percolation threshold of a network is a computationally expensive process which makes it necessary to have estimators that do not rely on numerical simulation. For the computation of the percolation threshold, there are several methods which do not involve numerical simulation of the process.

\subsection{Related Works}
In the past decade, the study of percolation in complex networks has gained a lot of attention. { The response of complex networks is investigated when subjected to attacks on vertices and edges for existing complex network models and real-world networks of scientific collaborations and Internet traffic \cite{perctimp}.} The network performance is quantitatively measured by the average inverse geodesic distance and the size of the largest connected cluster. A solution for the percolation of both single and interdependent networks of the hierarchical community structure of many networks is developed \cite{percolcn2}.

A k-leaf removal algorithm is introduced as a generalization of the so-called leaf removal algorithm in \cite{percolcn4}. The stability of the network's giant connected component is studied when facing any adverse events and modelled using link percolation \cite{percolcn5}. A detailed descriptive study that gives identical predictions with MPA using bond percolation as a benchmark is introduced \cite{percolcn6}. 

Percolation has also been studied in the context of interdependent and multiplex networks \cite{catastrophic,radicchiint,percolcn1,int1,int2,int3}. Moreover, it has been used to study various types of failures occurring in real-world networks \cite{explosive,group}. A variety of real-world systems like the spreading of diseases \cite{PhysRevE.62.7059}, electrical networks \cite{resistornet,GNN}, forest fires \cite{forestfires} and many other systems of importance \cite{impsys1, impsys2, impsys3, impsys4} have been studied under the percolation framework. It has also been used to study the effectiveness of the infrastructural and technological network, against random failure \cite{infracite}. 

Recent works have studied percolation in networks using the message passing approximation (MPA), which is known to provide accurate results on trees and inexact but good results, on real-world networks because of the existence of short loops \cite{PercSpars}. In \cite{percolcn3}, the message passing approximation was extended to account for the existence of clustering in networks, which yielded significant improvement when predicting the percolation diagram for site percolation. Recently, Cantwell et al. proposed a series of methods, considering the length of short loops in networks, taken during the message-passing approximation. These methods yield progressively better results on real-world networks \cite{mparecent}.

\subsection{Research Contributions}
In this paper, we use five machine learning regression models to estimate the percolation thresholds of bond, site and explosive percolation models represented through synthetic and real-world networks, by building a dataset consisting of 777 networks and training the models on this dataset. The mathematical model is built on sample data using machine learning algorithms for predicting or making decisions, without being precisely modelled for performing the task \cite{DLbook, MLbook}. 

The main contribution of this work is as follows:

\begin{itemize}
    \item Calculation of the true value of the percolation threshold is a computationally expensive process. Thus, there have been some empirical estimators of $\rho_c$. However, our proposed models have achieved better accuracy in predicting the bond, site and explosive percolation thresholds of various synthetic and real-world networks.
    \item In the case of bond percolation threshold, we have established a direct comparison between the performances of the empirical estimators and the machine learning models, with each model performing better. 
    \item We also evaluate the performances of each ML-based model. Of these, the gradient-boosting regressor and the random forest regressor perform the best.
    \item Finally we extend this model to predict the site and explosive percolation thresholds of these networks, which to the best of our knowledge, has rarely been explored previously.
\end{itemize}

The paper is organized as follows: a detailed explanation of the methodology with the mathematical description of the percolation threshold and a discussion on machine learning models are presented in section \ref{section:Method}. Section \ref{section:Materials} provides a description of the dataset and the features selected for regression analysis. In section \ref{section:Result} the detailed result analysis is presented with a discussion. Finally, section \ref{section:conclusion} concludes the paper.

\section{Methods}
\label{section:Method}
\subsection{Percolation Processes in Networks}

To demonstrate the efficacy and generality of the proposed approach, we consider three different percolation processes: bond, site and explosive percolation. The bond and site percolation models assume an underlying network structure, where the edges (or bonds) or nodes (or sites) are present independently with a probability $p$. At $p=0$, the network consists of nodes only at $p=1$ which provides the original network. At intermediate values of $p$, we find the networks in two phases: one where a large connected component exists and the other where numerous clusters of size much smaller than the size of the network exist. The probability value which splits the two phases is known as the percolation threshold ($\rho_c$). This value is an indicator of the robustness of the network in the occurrence of random failure \cite{perctimp}. It is therefore extremely important to know this value for a network in practical applications. 

More recently, there has been a lot of interest in the percolation transition's nature with a certain amount of intervention while adding the edges to a network. It was recently established that allowing these interventions have drastic and exciting consequences. In explosive percolation, when a new edge is to be added, two edges are chosen randomly and the edge resulting in the smallest giant connected component is added to the network. Such an intervention leads to a discontinuous percolation transition as opposed to the continuous transition observed in bond and site percolation processes \cite{d2015explosive}.

\subsection{Mathematical Description}
\subsubsection{Percolation Framework}
{ To compute the true value of percolation threshold ($\rho_c$), the following procedure is adopted \cite{PercSpars}: Given a connected, unweighted and undirected network with $n$ nodes and $e$ edges, we initialize the network with no edges and add edges sequentially in random order. The fraction of edges that have been added to the network is given by $p$. For a given value of $p$, the size of the second largest connected component \cite{MARGOLINA198273} is given by $S(p)$. We monitor the evolution of $S(p)$ concerning $p$ when the edges are sequentially added and the process is repeated $Q$ times yielding \cite{frad},

\begin{equation}
    P_\infty (p) = \frac{1}{nQ} \sum_{q=1}^{Q} S_q(p)
\end{equation}

Then, we compute susceptibility as:
\begin {equation}
 	\chi(p)=  \frac{(1/n^2 Q)\sum_{q=1}^{Q} (S_q(p))^2-[P_\infty (p)]^2} {P_\infty (p)} 	
\end {equation}

Finally, the best estimate of the percolation threshold ($\rho_c$) is the value of $p$ where the susceptibility reaches its maximum. \newline
\noindent Percolation Threshold, 
\begin {equation}
    \rho_c=arg{[\max \chi(p)]}
\end {equation}
}
However, this is not the only method to compute the true percolation threshold \cite{resiliency, GNN}. Another approach to compute $\rho_c$ is proposed by Newman and Ziff i.e., Monte-Carlo method \cite{EfficientMonteCarloAlgorithm}. The value obtained by the two methods is generally not the same, but for this paper, the difference between the obtained values is not significant. 

\subsubsection{Percolation Threshold Indicators} 
We consider three existing indicators of percolation threshold for comparison. The first estimator is the expected value of the percolation threshold in uncorrelated degree distribution with a given degree sequence \cite{P1a, P1b}.
\begin{align}
    P_1 = \frac{\langle k\rangle}{\langle k^2 \rangle-\langle k\rangle}
\end{align}
where $\langle k\rangle$ and $\langle k^2\rangle$ are the first and the second moments of the degree distribution of the graph, respectively.

The second and third estimators are the inverse of the largest eigenvalues of the adjacency matrix $A$ \cite{P2} and the non-backtracking matrix $M$ \cite{PercSpars}, respectively. ${\vec{v}}$ is the corresponding eigenvector. It is to be noted that, estimators $P_2$ and $P_3$ give a lower bound for the value of the percolation threshold. However, for dense networks, the difference between the true value of the percolation threshold and $P_2$ and $P_3$ is known to be small, and thus, these lower bounds are considered to be estimators of $\rho_c$ for dense networks.

\begin{align}
    P_2 = \left[\max _{\overline{v}} \frac{\vec{v}^{T} A \vec{v}}{\vec{v}^{T} \vec{v}}\right]^{-1}
\end{align}

\begin{align}
    P_3 = \left[\max _{\overline{v}} \frac{\vec{v}^{T} M \vec{v}}{\vec{v}^{T} \vec{v}}\right]^{-1}
\end{align}
Here, $M$ is the non-backtracking matrix defined as,
\begin{align}
    M=\left(\begin{array}{ll}{A} & {I-D} \\ I & {\emptyset}\end{array}\right)
\end{align}

In equation (6), $I$ is the identity matrix of size equal to the size of the corresponding network and $D$ is a diagonal matrix whose
elements are equal to the degrees of the nodes. The performance of these estimators has been investigated extensively in \cite{frad}.

\subsection{Machine Learning Models}
In this paper, we utilise five predictive machine learning methods, namely Linear Regression, Artificial Neural Network (ANN), Random Forest Regression, Multilayer Perceptron-based (MLP) Regression (modified feed-forward ANN) and Gradient Boosting Regression (GBR).

In the case of regression, a machine learning model attempts to learn the relation between the features of the dataset and the output attribute. In this case, structural and statistical properties of networks act as the features of the dataset and the output attribute is the numerically computed percolation threshold of the network, which the machine learning models attempt to predict.

\subsubsection{Linear Regression}
Linear Regression is an elementary machine learning model, where the model trains to find the linear relationships between the features and the output attribute \cite{MLbook, SLbook}. Here a slightly advanced linear regression model is used, which also models the linear relationship between the basic non-linear functions (like polynomial, exponential, logarithmic, etc.) of the features. Then, the parameters of the model, i.e., the coefficients of the features in the linear representation are optimised to best fit the training data. 

From the Scikit-learn toolkit, the linear regression model consists of coefficients $w = (w_1, ..., w_p)$ to minimize the difference of the sum of squares between the input features of the dataset, $X$, and the output attributes predicted by the model, $y$. The linear regression model is described in function form below.
\begin{equation}
    \min_{w} || X w - y||_2^2
\end{equation}

\begin{figure*}[t!]                                                
\centering
    \begin{center}                                                 
        \includegraphics[width=1.5\columnwidth]{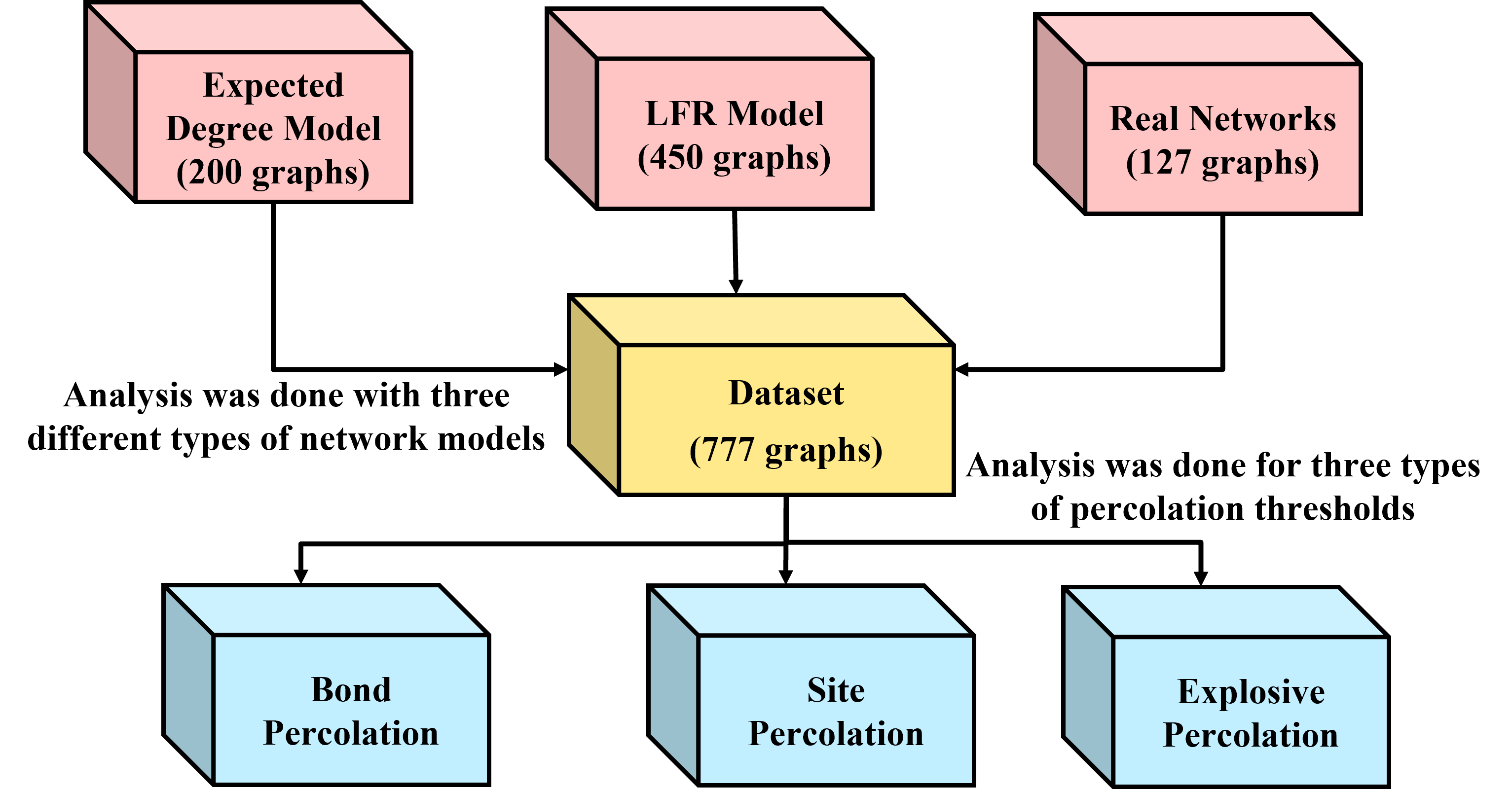}
        \caption{Illustrates the types of network models utilised, and the number of networks generated using each model, to form the total dataset. This dataset is analysed for three different types of percolation processes, i.e., bond, site and explosive percolation}                    
        \label{fig:dataset}                                
    \end{center}                                                   
\end{figure*} 
\subsubsection{Simple Artificial Neural Network}
An Artificial Neural Network (ANN) is a predictive model resembling biological neural networks. It attempts to learn the complex non-linear relationships between the features and the output attribute \cite{MLbook}. This is done by computing multiple non-linear representations of the input features that are essential for the accurate prediction of the output. A key point of difference it has with logistic regression is that ANNs have one or more non-linear layers between the input and the output layer. These intermediate non-linear layers are called hidden layers. { The non-linearity is introduced by the activation function, we have used Rectified Linear Unit (ReLU) in the analysis}. Mathematically, for a single-layer ANN with $m$ inputs, we represent it as:
\begin{equation}
    y_{input} = \sum_{i}^{m}x_{i}.w_{i}
\end{equation}
where the inputs are $x_{1}$, $x_{2}$, ..., $x_m$, having weights associated with them $w_{1}$, $w_{2}$, ..., $w_m$, respectively.\\
The output of the network is found using an activation function, $F$,
\begin{equation}
    Y_{output} = F(y_{input})
\end{equation}
For our experiments, we are using a simple single-layered ANN, having 16 nodes.

\subsubsection{Random Forest Regression}
The Random Forests model is an ensemble of decision trees, which trains a large number of trees with randomly chosen subsets of features to predict the output attribute. These trees then predict the output independently, and the average of all these outputs is taken, to obtain a better prediction. \cite{rf}.

Random Forest Regression models are quite robust and generally perform great on a variety of mathematically intensive problems, especially on features with non-linear relationships. However, the number of trees to include in the model must be carefully assigned as it is susceptible to over-fitting.
In a typical Random Forest model, we begin with $k$ arbitrary data points taken from the training set and form a decision tree corresponding to these $k$ data points.
We then assign $N$ number of trees, necessary for the task and again continue the previous steps.
For a new data point, each tree from the set of $N$ decision trees predicts the value of $y$ for the given data point considered and correlates the newly selected data point to the mean of all of the predicted $y$ values.
\\Mathematically, for each decision tree, the Scikit-learn model uses Gini Importance \cite{menze2009comparison} to compute the importance of a node, assuming only two child nodes originate from each parent node (binary tree scheme):
\begin{equation}
    ni_{j} = w_{j}C_{j} -  w_{left(j)}C_{left(j)} - w_{right(j)}C_{right(j)}
\end{equation}
where $ni_{j}$ is the node importance of the particular node $j$, 
$w_{j}$ is the weighted number of samples arriving at node $j$, 
$C_{j}$ refers to the impurity value of node $j$, $left(j)$ and $right(j)$ are the child nodes from the left split and the right split on node $j$, respectively. This expression is used to calculate the feature importance (discussed in later sections) at each tree level, and also at the final Random Forest level.

\subsubsection{Multilayer Perceptron}
Multi-layer Perceptron (MLP) is a supervised learning algorithm which learns a general function $f(.):R^{m}\longrightarrow R^{o}$ by undergoing training on a given dataset. Here $m$ is the number of dimensions for the input features and $o$ is the number of dimensions for the output features. When we have a set of input features $X = x_{1}, x_{2},.., x_{m}$ and a target $y$ to be predicted, the MLP can learn from a non-linear function approximator for both kinds of prediction assignments: classification or regression. It is a multilayer feed-forward ANN.

MLP has salient advantages such as real-time learning using \texttt{partial\_fit}, and the capacity to learn non-linear models. However, it falls short to feature scaling and needs tuning of multiple parameters.
Class \texttt{MLPRegressor} realizes a multi-layer perceptron (MLP) which trains using backpropagation having no activation function present in the output layer. The output layer is interpreted to utilize the identity function as an activation function. Hence, the square error is used as the loss function, and the output is a set of continuous values.

\subsubsection{Gradient Boosting Regression}
Gradient Boosting machines are a family of potent machine learning algorithms that have achieved notable success in a variety of real-world applications. The forward stage-wise additive model construction used by a Gradient Boosting regressor enables the optimization of any arbitrary differentiable loss function. A regression tree is fitted on the negative gradient of the provided loss function at each level \cite{gbr2}. They may be significantly tailored to the specific requirements of the application by learning about various loss functions \cite{GBR}. 
We have set the parameter \texttt{n\_estimators} at 1000, which is the number of boosting stages to perform. A big number typically results in greater performance since gradient boosting is highly resistant to over-fitting. \texttt{Learning\ rate} is kept default since there is a trade-off with \texttt{n\_estimators}.   

\section{Materials}
\label{section:Materials}
% \textit{tools, dataset, apparatus used. preprocessing, explanation of filtering, choosing from repository}
\subsection{Dataset}
The performance of a machine learning model is strongly dependent on the quality of data used to train the algorithm. Therefore, the synthetic networks used to train the models (in addition to the real-world networks) must be as close to the real-world networks as possible. A series of models have been proposed to account for the resultant evolution of real-world networks. 

The dependence of the structural properties of networks on percolation has been widely studied. For example, if the clustering coefficient assumes a large value, it is known to give rise to networks with a lower value of $\rho_c$ and a decreased size of the giant component of the network. Moreover,  it also influences the spreading processes on networks. \cite{clustered1,clustered2,clustered3}.

\renewcommand{\arraystretch}{1.12}
\begin{table*}
\caption{Basic Statistics of the Dataset}
\label{table:basicstk}
\centering
    \begin{tabularx}{\textwidth}
    {
        | >{\centering\arraybackslash\hsize=0.125\hsize}X 
        | >{\centering\arraybackslash\hsize=0.125\hsize}X
        | >{\centering\arraybackslash\hsize=0.125\hsize}X
        | >{\centering\arraybackslash\hsize=0.125\hsize}X
        | >{\centering\arraybackslash\hsize=0.125\hsize}X
        | >{\centering\arraybackslash\hsize=0.125\hsize}X 
        | >{\centering\arraybackslash\hsize=0.125\hsize}X 
        | >{\centering\arraybackslash\hsize=0.125\hsize}X |
    }
        \hline
        \textbf{Type} & \textbf{Minimum} & \textbf{Maximum} & \textbf{Mean} & \textbf{Median} & \textbf{Standard Deviation} & \textbf{Skewness} & \textbf{Kurtosis}\\
        \hline
        Bond & 0.0089 & 0.3184 & 0.1053 & 0.0669 & 0.0923 & 0.5939 & -1.1661\\
        Site & 0.0188 &	0.4838 & 0.1318 & 0.1112 & 0.0909 & 0.4318 & -1.0103\\
        Explosive & 0.0206 & 0.4852 & 0.2083 & 0.1604 & 0.1627 & 0.2973 & -1.5631\\
        \hline
    \end{tabularx}
\end{table*}

 \begin{table}[h]
\caption{Range of parameters, from which linearly distributed values are used for synthesizing Expected degree model networks}
\label{table:tab-ConfigParams}
\centering
    \begin{tabular}{|c|c|}
        \hline
        \textbf{Parameters} & \textbf{Range of values}\\
        \hline
        $alpha$ & 20 values in [2, 4]\\
        $n$ & 10 values in [1000, 5000]\\
        \hline
    \end{tabular}
\end{table}

To generate synthetic networks, we use the network evolution model by Ghoshal et al. \cite{final}. It uses various parameters to account for different elementary processes, which generate networks that are closer to the real world both structurally and statistically. It also provides variability in the obtained data. 
We select the parameters of the model based on several conditions to obtain 777 networks with sizes ranging from $10^3$ to $5 \times 10^3$ as shown in FIG. \ref{fig:dataset}. In addition to this, we use various social, technological, biological, transportation, citation, economic, and other miscellaneous networks. A total of 300 real-world networks from the Network Repository \cite{nr-aaai15} of sizes ranging from $1000$ to $1.5\times10^4$ have been included in the generated dataset. The statistical information of the dataset is enlisted in TABLE \ref{table:basicstk}.

The true value of the percolation threshold is computed for $Q=10000$, using numerical simulations through the Dirac Supercomputing facility which is 78.8 Teraflops and includes 60 Teraflops of CPU and 4$\times$4.7 Teraflops of GPU. We run the numerical computation of the percolation threshold using 48 core parallel processing node of Dirac. This calculated percolation threshold value is the output feature of the dataset. To train the machine learning models, one needs to select the structural and statistical properties of networks that may affect the percolation threshold of the networks as features. We consider the following properties of the networks as features: Average degree, minimum degree, clustering coefficient, degree assortativity, and power-law degree distribution. We calculated these features on Intel Core i5-10210U devices with CPU including 1.60GHz.

%edit:configuration models -> expected degree model
%edit:remove preprocess subsection
We have prepared a dataset of two types of synthetic networks, namely the Expected degree model and LFR model, along with real network templates. While preparing the expected degree models, it was based on several parameters, enlisted below in TABLE \ref{table:tab-ConfigParams}.

\begin{table}[h!]
\caption{Range of parameters, from which linearly distributed values are used for synthesizing LFR Models}
\label{table:tab-LFRParams}
\centering
    \begin{tabular}{|c|c|}
        \hline
        \textbf{Parameters} & \textbf{Range}\\
        \hline
        $n$ & 2500\\
        $mu$ & 3 values in [0.1, 0.9]\\
        $\tau_1$ & 9 values in [2, 4]\\
        $\tau_2$ & 3 values in [3, 5]\\
        $maxc$ & 500\\
        $minc$ & 40\\
        $k$ & 5 values in [10, 50]\\
        $max\_k$ & 500\\
        \hline
    \end{tabular}
\end{table}

\textbf{Expected Degree Model}: 
This model is based on an algorithm that utilises an $n$-length long sequence of expected degrees $W=(w_0,w_1,...,w_{n-1})$. It correlates an edge of the network between two of its nodes, $u$ and $v$ with the probability,
\begin{equation}
    p_{uv}=\frac{w_{u}w_{v}}{\Sigma_k w_k}
\end{equation}
The complexity of the expected degree model-generating algorithm is $O(n+m)$, where $n$ denotes the number of nodes and $m$ denotes the expected number of edges \cite{Chung2002}. Since randomly varied distributions used for alpha, may lead to skewed or biased results, we have chosen to use a range of uniformly distributed values of alpha.

\textbf{LFR Model}: Lancichinetti–Fortunato–Radicchi benchmark is an algorithm which helps in synthesizing artificial networks that look like real-world networks. They have already established communities and utilized them for comparison using various community detection methods. These generated benchmark models are found effective as it exposits for the heterogeneous distributions of the node degrees and community size \cite{shen2013community}. While preparing the LFR, it was based on several parameters, enlisted below in TABLE \ref{table:tab-LFRParams}.

% We also undertake some preprocessing steps to ensure proper performance by the dataset.

% \subsection{Filtering}
% After preprocessing, we checked the dataset for any abnormal values of percolation thresholds or other features. A few graphs that behaved as outliers were thus discarded, since including them could affect the training process of the predictive models.

\subsection{Feature Importance}
%-------------------------------------------------------------------------

The features selected for the regression analysis of the percolation threshold are known to affect the structural properties of the networks being used in the dataset. 
Five features are being used for the learning process of the machine learning models, to yield bond, site and explosive percolation thresholds. These features are:
\begin{itemize}
    \item Average Degree: It is defined as the average number of edges per node in the graph. It is given by:
    \begin{equation}
        Average\ Degree=\frac{Total\ number\ of\ edges\ (E)}{Total\ number\ of\ nodes\ (N)}
    \end{equation}
    \item Minimum Degree, $\delta(G)$: It is denoted as the minimum number of edges incident on a node, in a graph, { whose removal will produce a disconnected graph} $G$ \cite{MD}.
    \item Clustering Coefficient: It is a mathematical estimation of the extent to which the nodes in a given graph tend to cluster together. A large value of the clustering coefficient is known to give rise to networks with a lower value of $\rho_c$ and a decreased size of the giant component of the network. Moreover, it also influences the spreading processes on networks \cite{CC}.

    \begin{figure}[h!]
\begin{subfigure}{0.35\textwidth}
    \includegraphics[width=\textwidth]{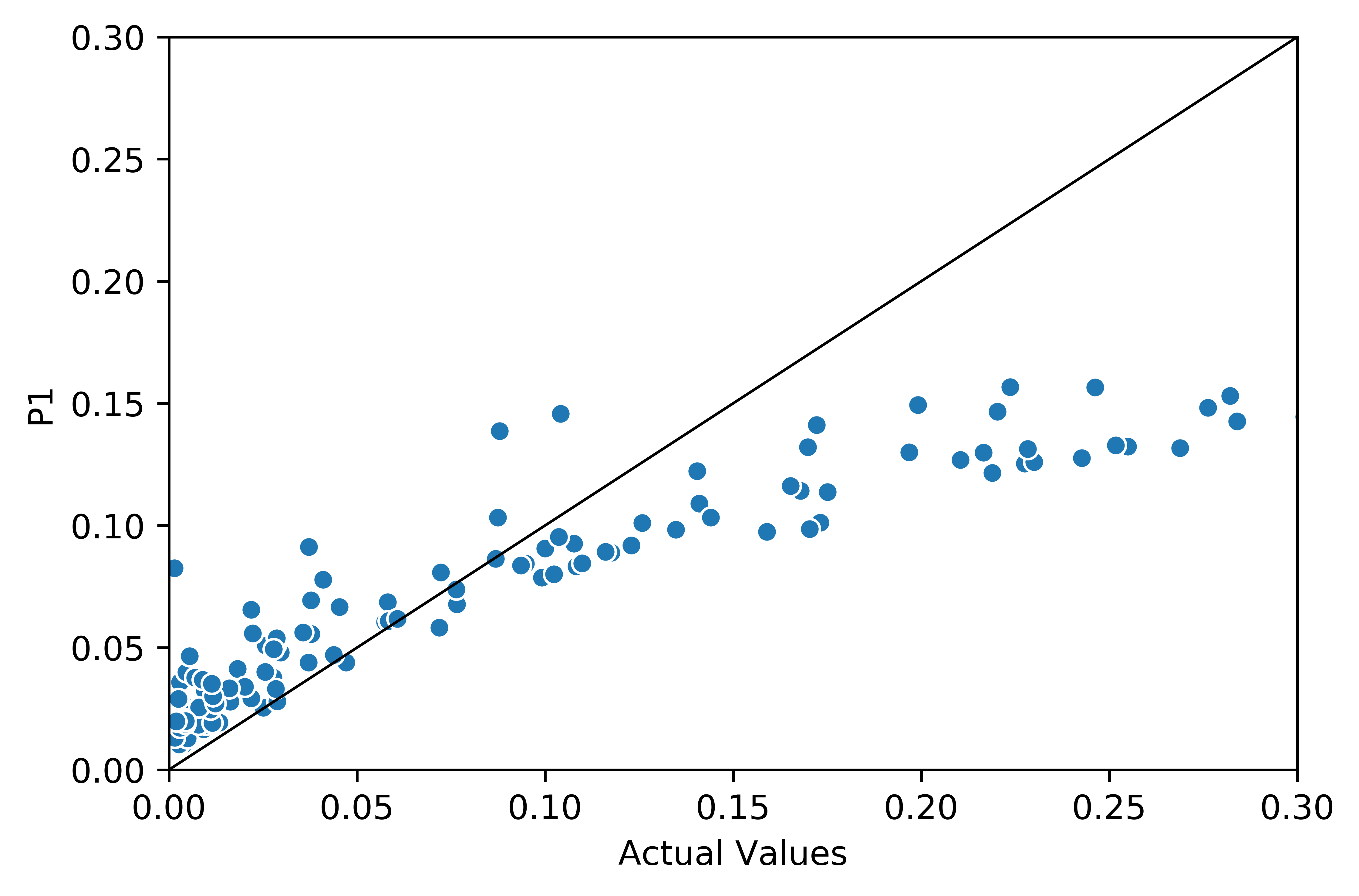}
    \caption{$P_1$}
    \label{fig:p1}
\end{subfigure}
\hfill 
\begin{subfigure}{0.35\textwidth}
    \includegraphics[width=\textwidth]{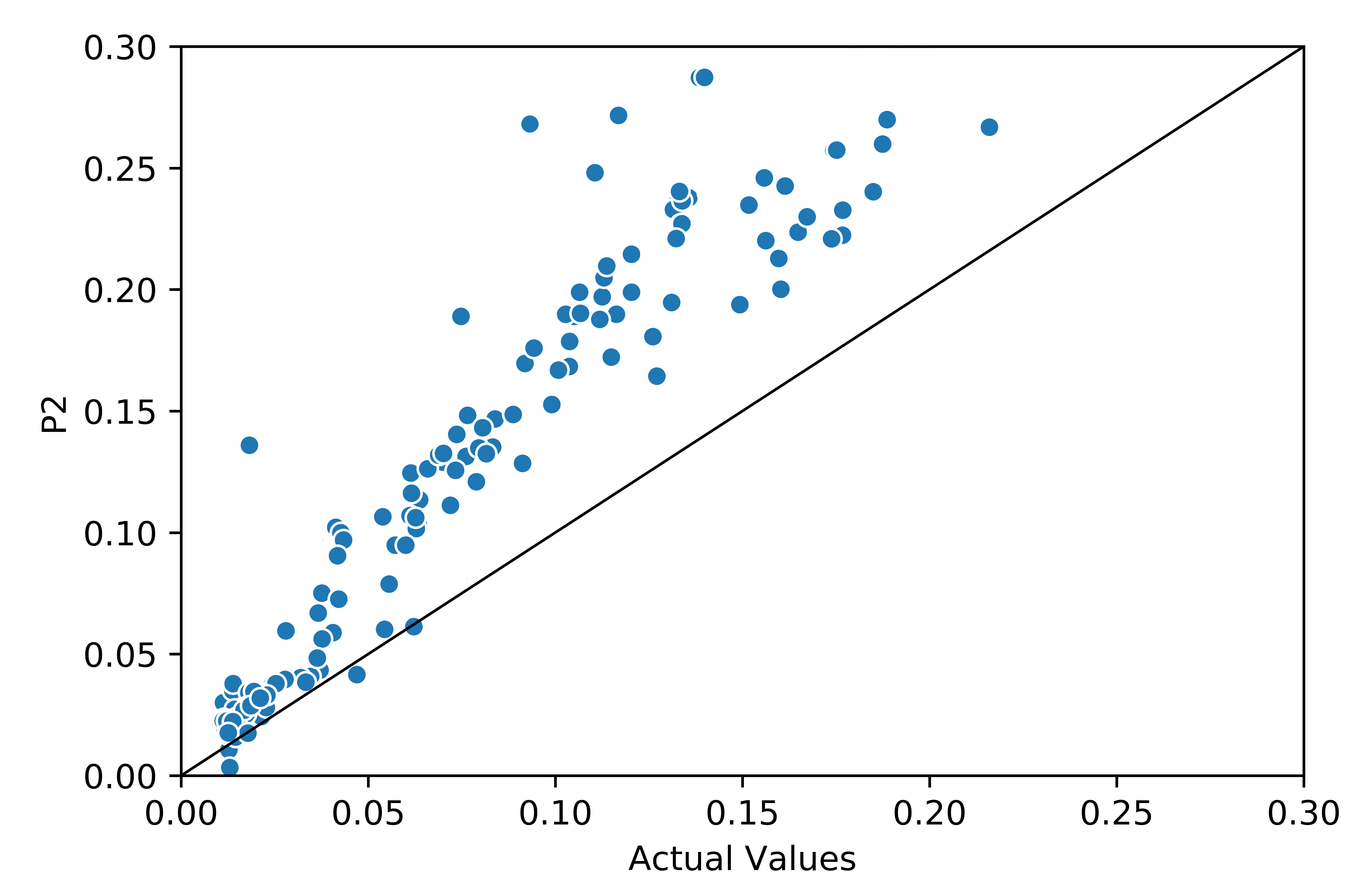}
    \caption{$P_2$}
    \label{fig:p2}
\end{subfigure}
\hfill
\begin{subfigure}{0.35\textwidth}
    \includegraphics[width=\textwidth]{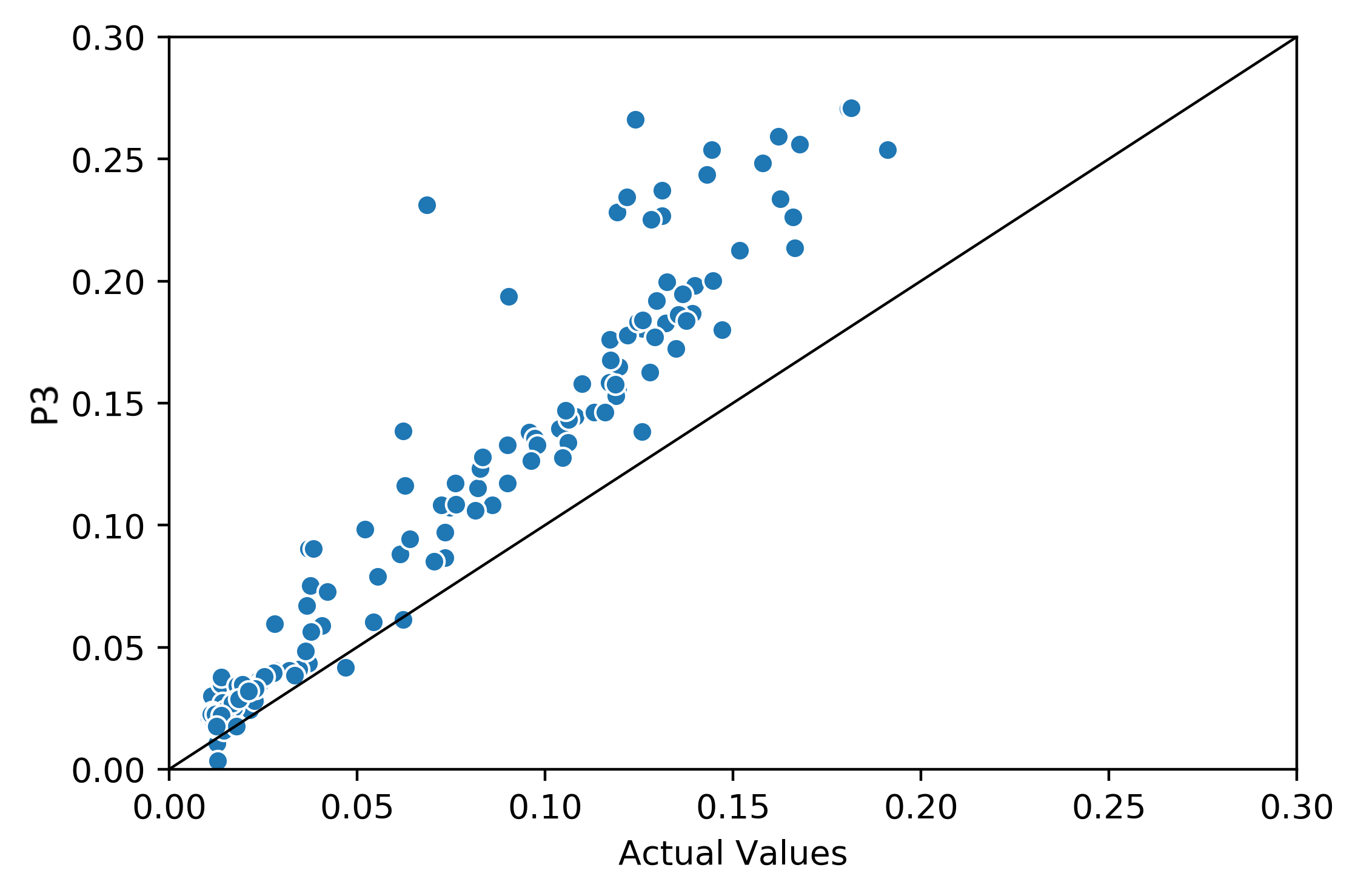}
    \caption{$P_3$}
    \label{fig:p3}
\end{subfigure}
\caption{Q-Q plots between predicted and actual values for the empirical estimators of $P_1$,$P_2$ and $P_3$}
\label{fig:estimators}
\end{figure}

\begin{table}[h!]
\caption{RMSE measures for the three empirical estimators calculated for bond percolation threshold}
\label{table:tab-empirical}
\centering
    \begin{tabular}{|c|c|}
        \hline
        \textbf{Predictor} & \textbf{RMSE}\\
        \hline
        $P_1$ & 0.276084\\
        $P_2$ & 0.171556\\
        $P_3$ & 0.135102\\
        \hline
    \end{tabular}
\end{table}

% BOND PERCOLATION THRESHOLD RESULTS
\begin{figure*}[t!]
\begin{center}
\begin{subfigure}{0.325\textwidth}
    \includegraphics[width=\textwidth]{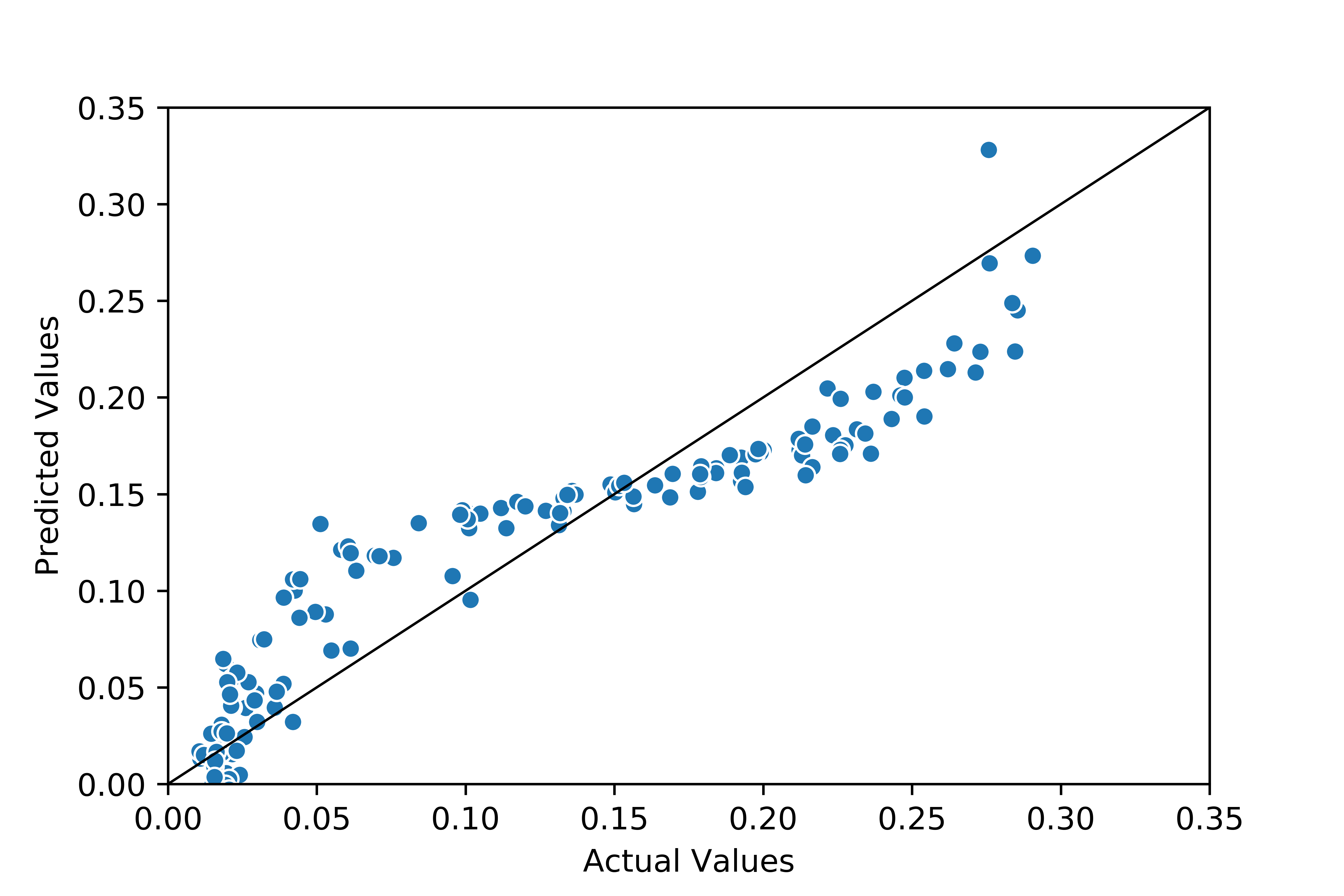}
    \caption{Linear Regression}
    \label{fig:BondLR}
\end{subfigure}
\hfill
\begin{subfigure}{0.325\textwidth}
    \includegraphics[width=\textwidth]{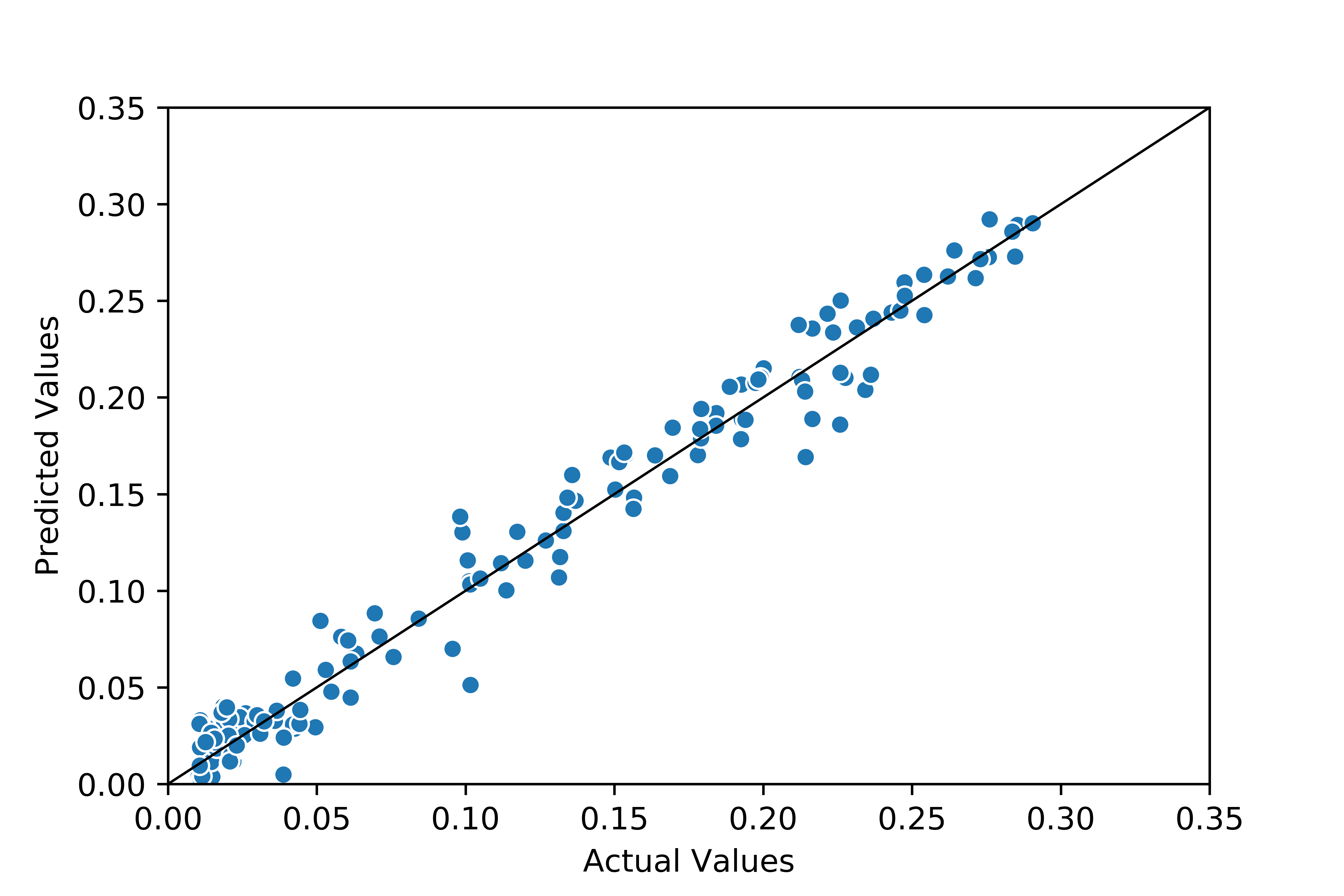}
    \caption{Simple ANN}
    \label{fig:BondANN}
\end{subfigure}
\hfill
\begin{subfigure}{0.325\textwidth}
    \includegraphics[width=\textwidth]{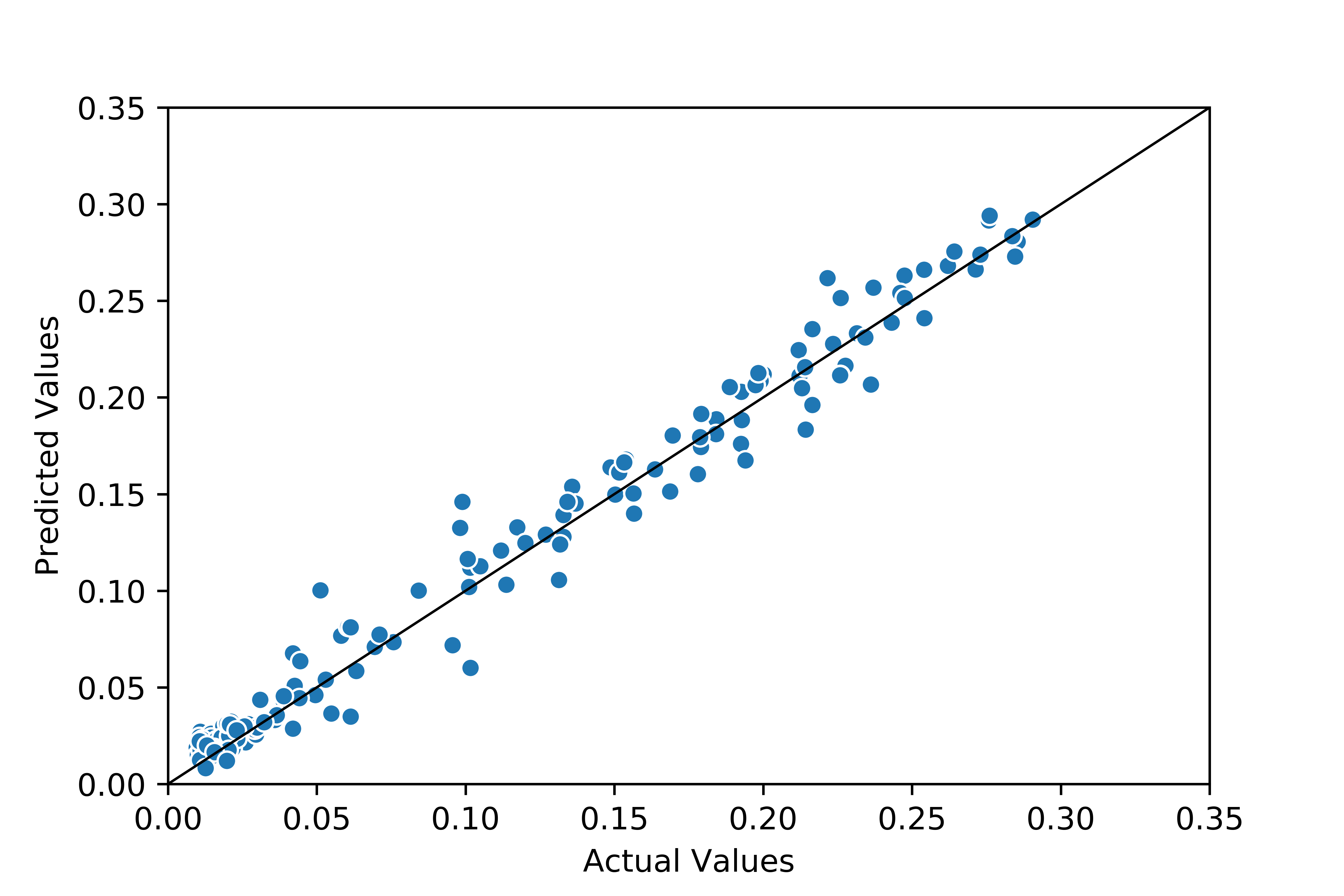}
    \caption{Random Forests Regressors}
    \label{fig:BondRFR}
\end{subfigure}
\hfill
\begin{subfigure}{0.325\textwidth}
    \hspace{15cm}
    \includegraphics[width=\textwidth]{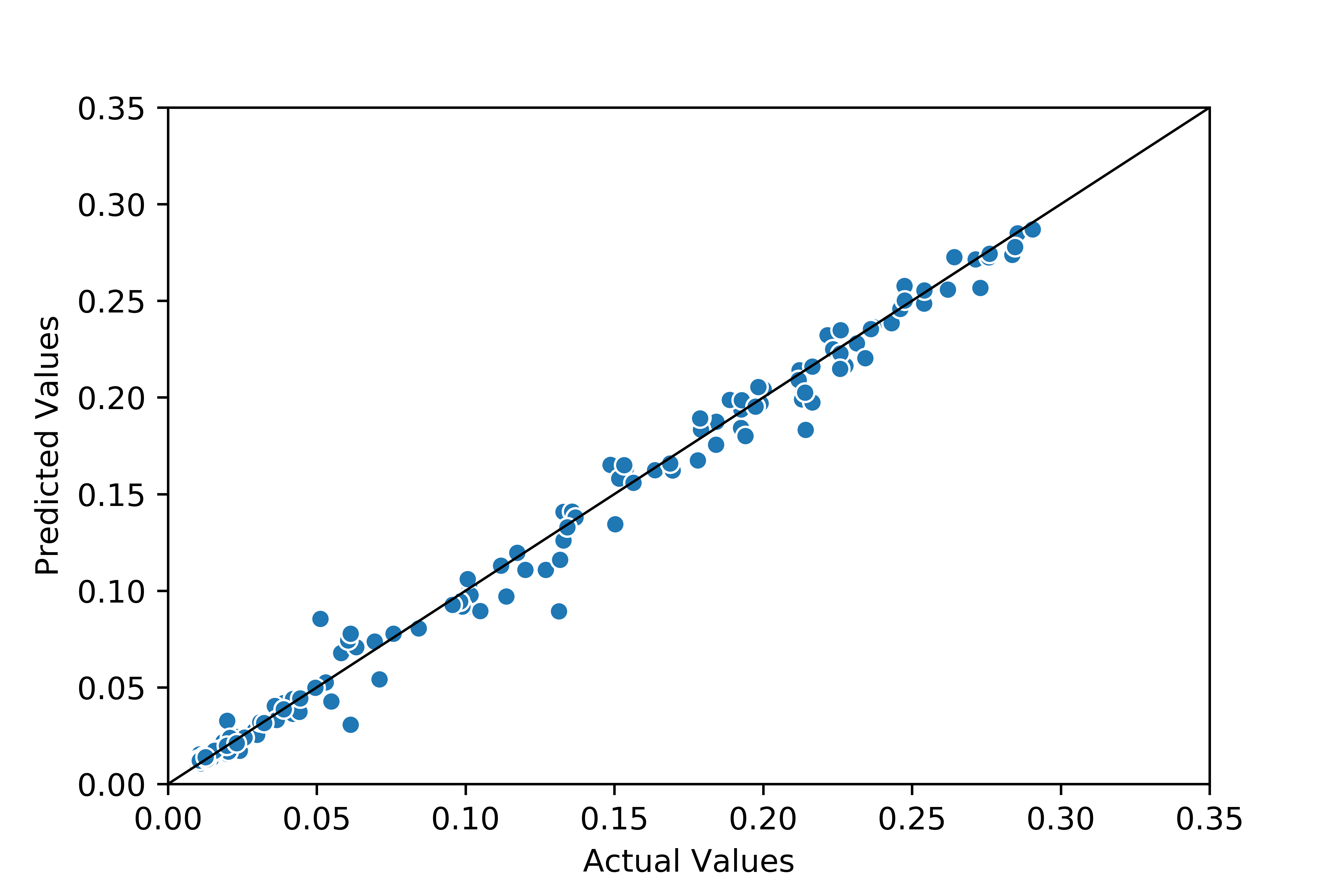}
    \caption{Multilayer Perceptron}
    \label{fig:BondMLP}
\end{subfigure}
\begin{subfigure}{0.325\textwidth}
    \includegraphics[width=\textwidth]{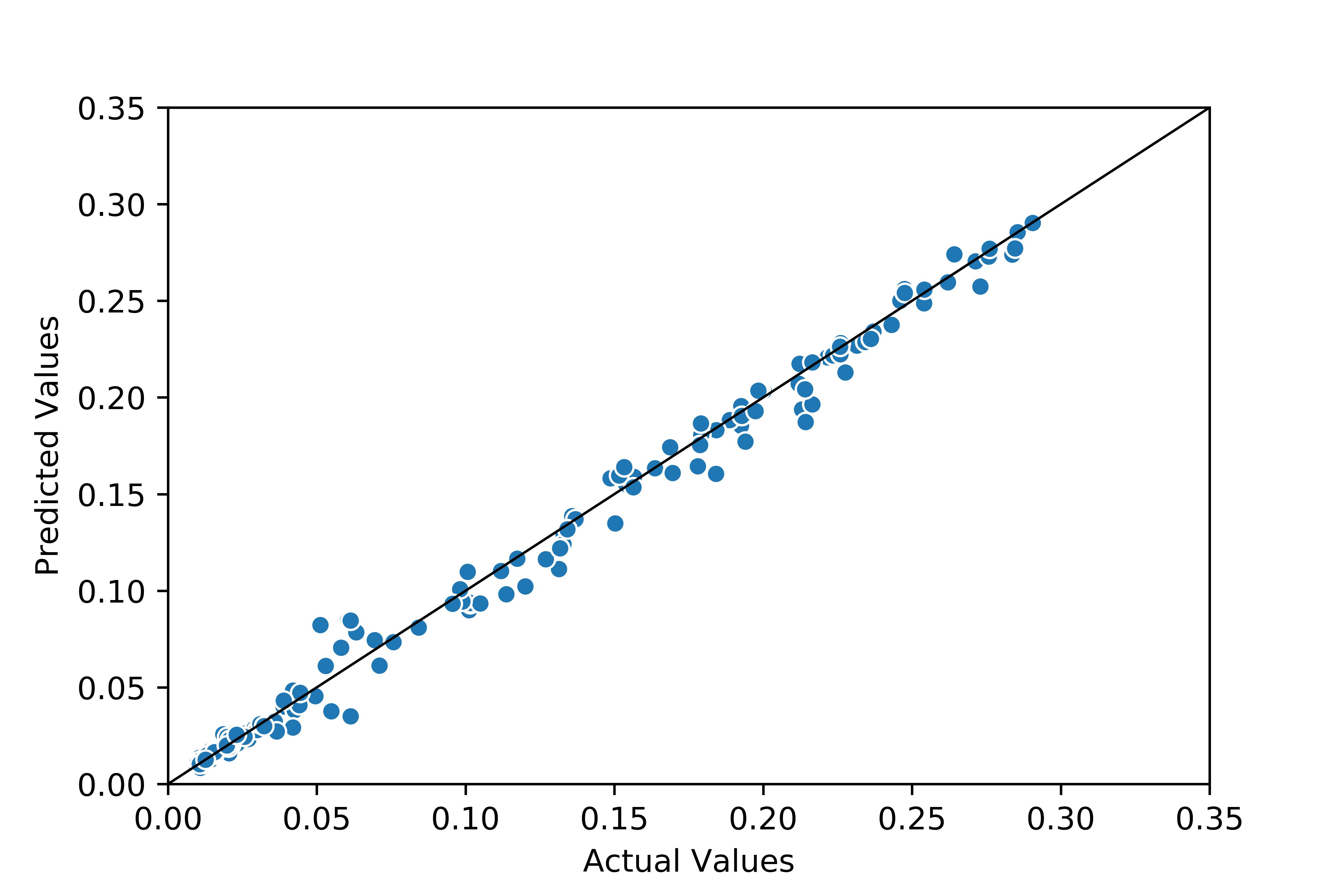}
    \caption{Gradient Boosting Regressor}
    \label{fig:BondGBR}
\end{subfigure}
\end{center}       
\caption{Performance of the machine learning models in predicting bond percolation threshold, as plotted with predicted values in the y-axis and actual values in the x-axis}
\label{fig:Bond}
\end{figure*}

\renewcommand{\arraystretch}{1.12}
\begin{table*}[t!]
\caption{Comparison of the root mean squared errors (RMSE) for the various methods with or without Feature importance selection}
\label{table:results}
\centering
    \begin{tabularx}{\textwidth}
    {
        | >{\centering\arraybackslash\hsize=0.11\hsize}X 
        | >{\centering\arraybackslash\hsize=0.11\hsize}X
        | >{\centering\arraybackslash\hsize=0.11\hsize}X
        | >{\centering\arraybackslash\hsize=0.11\hsize}X
        | >{\centering\arraybackslash\hsize=0.11\hsize}X
        | >{\centering\arraybackslash\hsize=0.11\hsize}X 
        | >{\centering\arraybackslash\hsize=0.11\hsize}X 
        | >{\centering\arraybackslash\hsize=0.11\hsize}X
        | >{\centering\arraybackslash\hsize=0.11\hsize}X |
    }
        \hline
        \textbf{$p_c$ type} & \multicolumn{2}{c|}{\textbf{LR}} & \multicolumn{2}{c|}{\textbf{ANN}} &  \multicolumn{2}{c|}{\textbf{RFR}} &  {\textbf{MLP}} &  {\textbf{GBR}} \\
        \cline{2-7}
         & \textbf{without} & \textbf{with} & \textbf{without} & \textbf{with} & \textbf{without} & \textbf{with} &  & \\
         \hline
        Bond & 0.038543 & 0.038331 & 0.027319 & 0.015919 & 0.008662 & 0.008652 & 0.008892 & 0.007756\\
        Site & 0.034628 & 0.034795 & 0.022097 & 0.017044 & 0.008665 & 0.008010 & 0.009166 & 0.008114 \\
        Expl & 0.066159 & 0.065667 & 0.020287 & 0.045706 & 0.008311 & 0.008200 & 0.008550 & 0.007334 \\
        \hline
    \end{tabularx}
\end{table*}

    \item Degree Assortativity: Assortativity is a mathematical estimation of the similarity of connections in the graph regarding the node degree. This result is explained in \cite{PhysRevE.67.026126}, and is given by,
    \begin{equation}
        r=\frac{\sum_{xy}^{} xy(e_{xy} - a_{x}b_{y})}{\sigma_a\sigma_b}
    \end{equation}
    where $e$ denotes the joint probability distribution or mixing matrix of the node degrees, $\sigma_a$ and $\sigma_b$ stand for the standard deviations of the distributions $a_x$ and $b_y$. The value of $r$ varies within $-1 \leq r \leq 1$, where $r=1$ denotes perfect assortativity and $r=-1$ denotes perfect disassortativity, which in turn means the negative correlation between x and y \cite{PhysRevE.67.026126}. 
   \item Power-law Degree Distribution: There are two main types of power-law distributions: discrete distributions, where the quantity of interest only assumes a discrete set of values, usually positive integers, and continuous distributions, which associate with continuous real numbers. Let the quantity $x$ whose distribution we are interested to represent. A probability density, $p(x)$, that is continuous is used to describe a power-law distribution such that,
\begin{equation}
        p(x)dx = Pr(x \leq X \leq x+dx) = Cx^{-\alpha}dx
    \end{equation}  where $X$ denotes the observed value, $C$ denotes a normalization constant \cite{Clauset_2009}.
\end{itemize}
We used Random Forest Regressor to calculate the feature importance as well and found that for all three types of the percolation threshold. They are estimated in the form of the mean and standard deviation of accumulation of the impurity decrement within each tree of the set of $N$ trees generated.

% ESTIMATORS RESULTS
%--------------------------------------------
% \begin{table}[h!]
% \caption{RMSE measures for the three empirical estimators calculated for bond percolation threshold}
% \label{table:tab-empirical}
% \centering
%     \begin{tabular}{|c|c|}
%         \hline
%         \textbf{Predictor} & \textbf{RMSE}\\
%         \hline
%         $P_1$ & 0.276084\\
%         $P_2$ & 0.171556\\
%         $P_3$ & 0.135102\\
%         \hline
%     \end{tabular}
% \end{table}

\section{Results and Discussion}
\label{section:Result}

%---------------------------------------------
% SITE PERCOLATION THRESHOLD RESULTS
\begin{figure*}[t!]
\centering
\begin{subfigure}{0.325\textwidth}
    \centering
    \includegraphics[width=\textwidth]{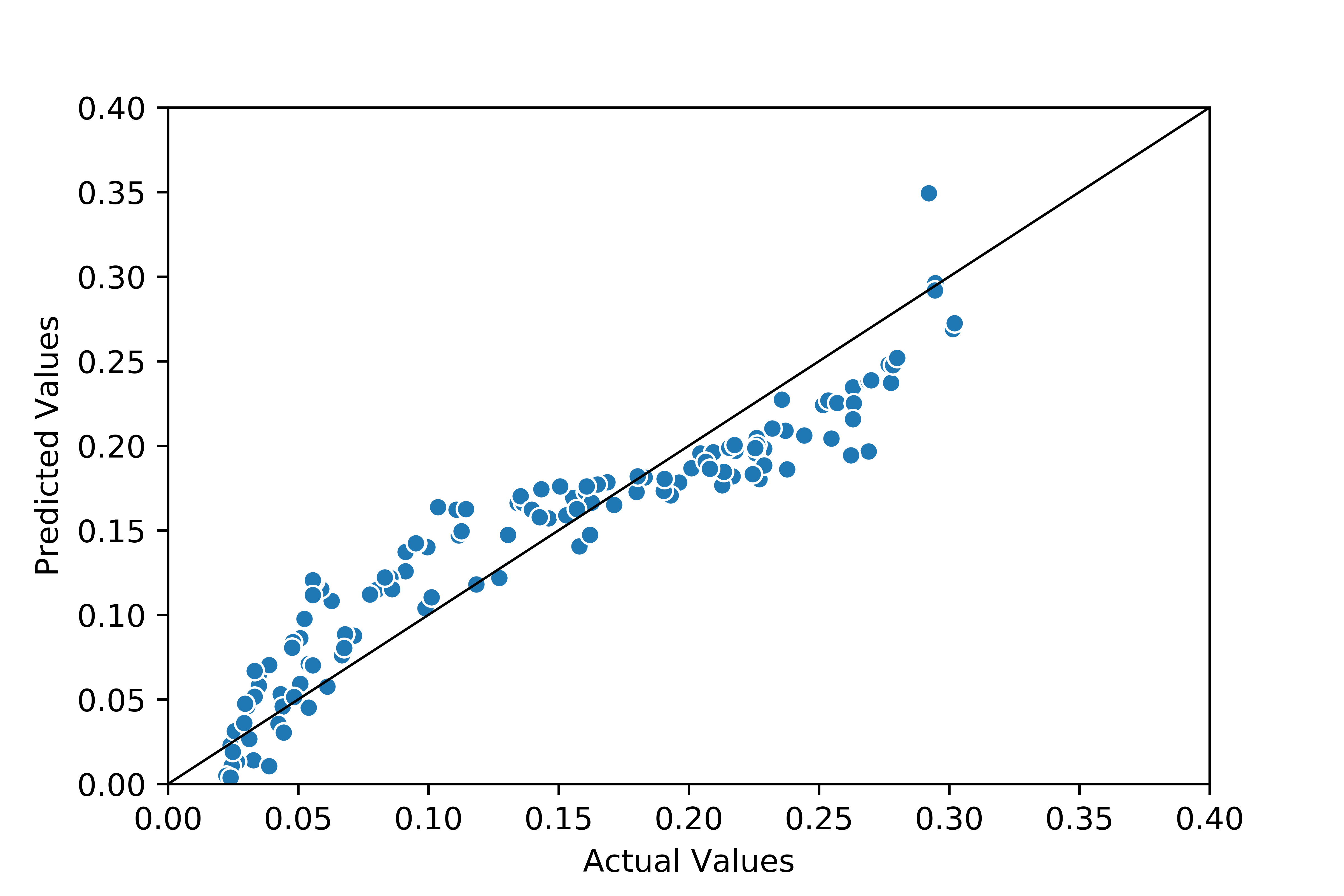}
    \caption{Linear Regression}
    \label{fig:SiteLR}
\end{subfigure}
\hfill
\begin{subfigure}{0.325\textwidth}
    \centering
    \includegraphics[width=\textwidth]{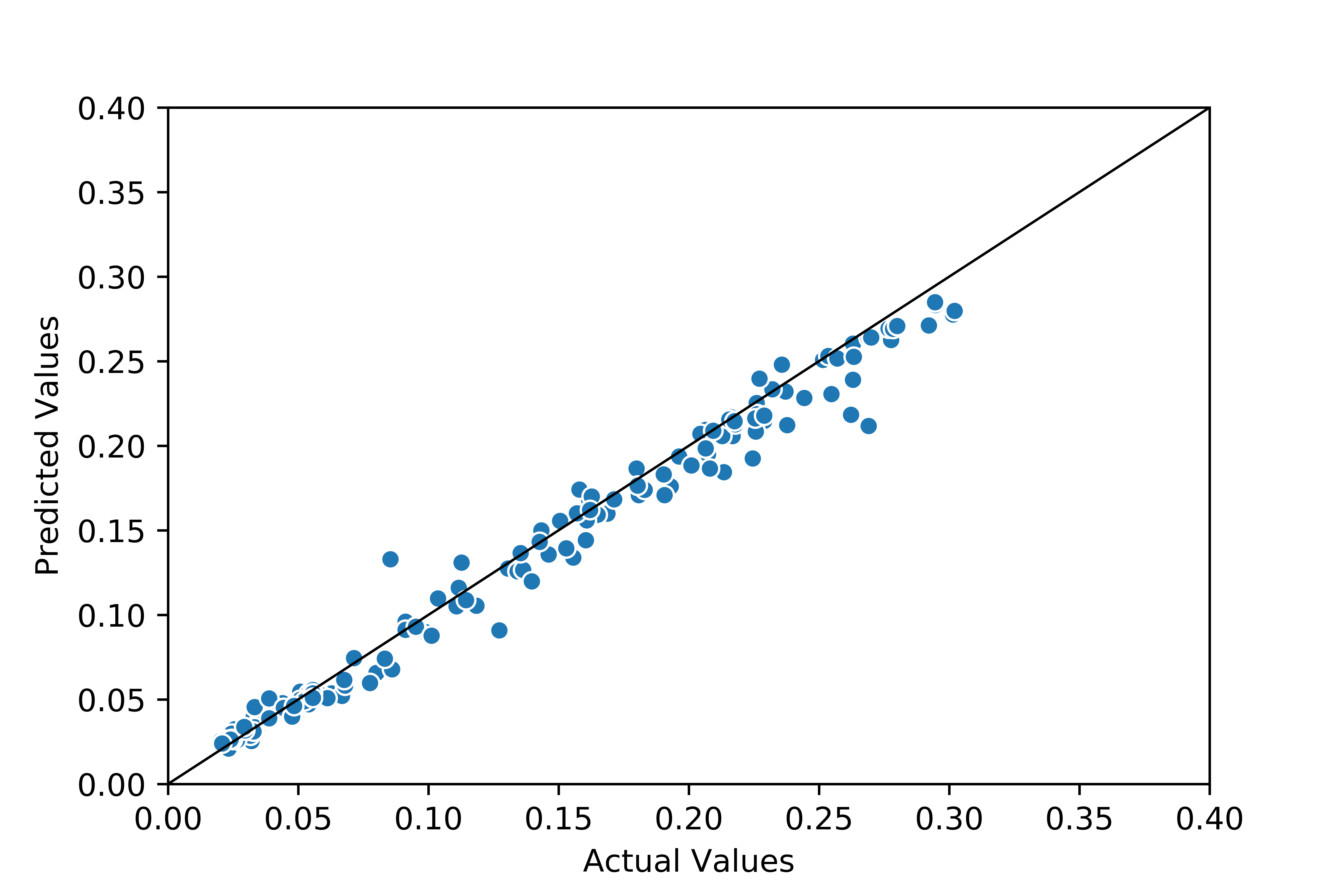}
    \caption{Simple ANN}
    \label{fig:SiteANN}
\end{subfigure}
\hfill
\begin{subfigure}{0.325\textwidth}
    \centering
    \includegraphics[width=\textwidth]{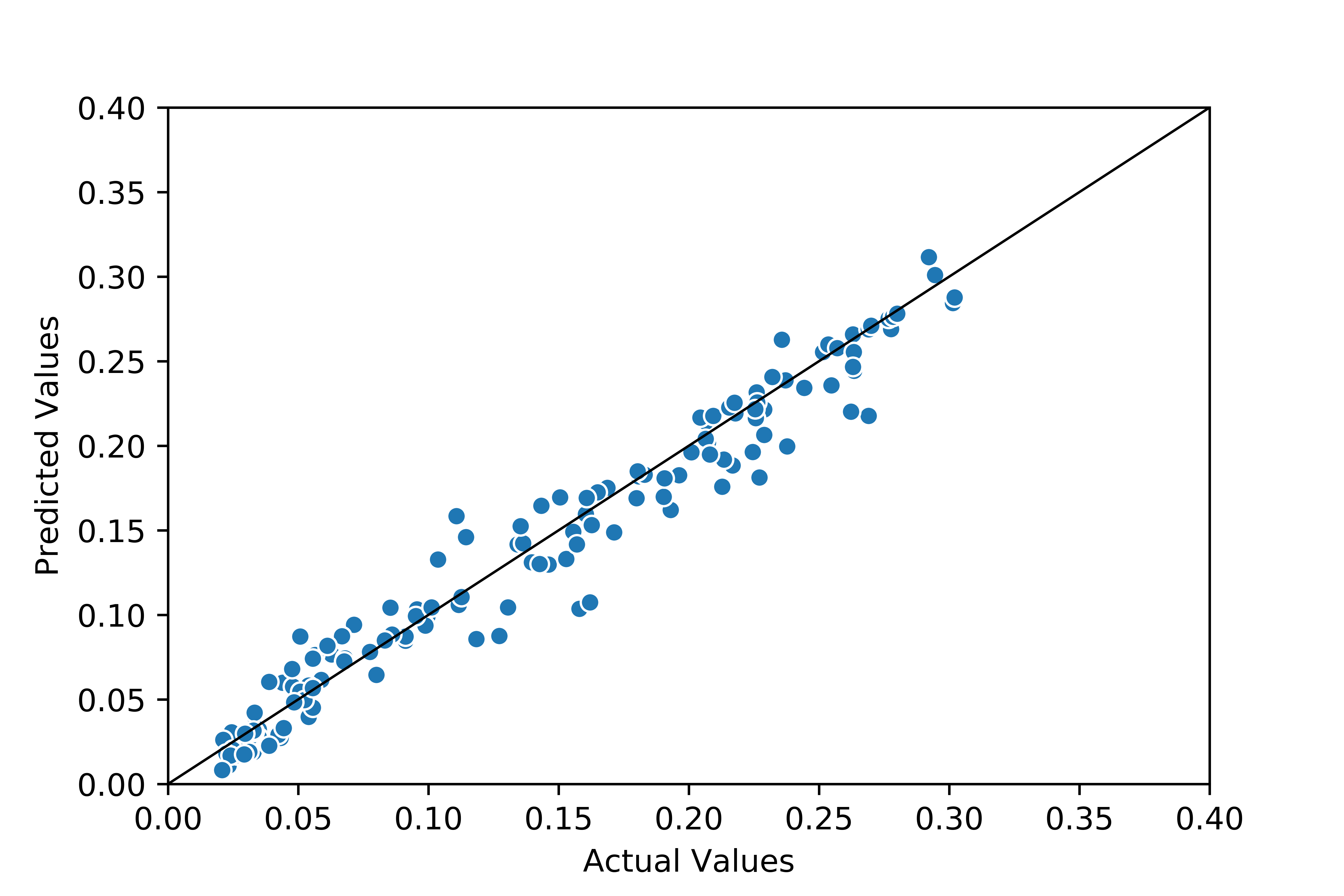}
    \caption{Random Forests Regressors}
    \label{fig:SiteRFR}
\end{subfigure}
\hfill
\begin{subfigure}{0.325\textwidth}
    \centering
    \includegraphics[width=\textwidth]{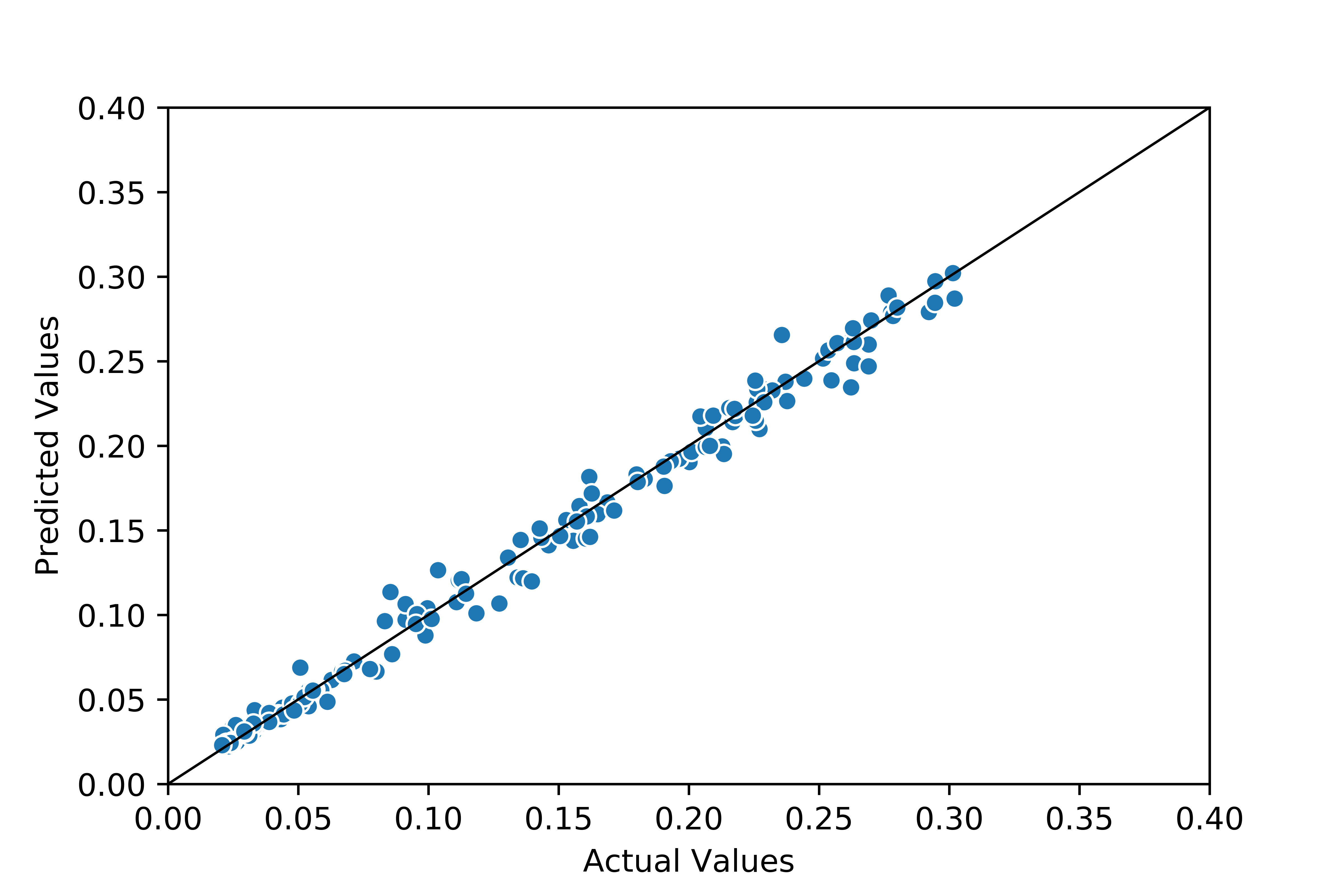}
    \caption{Multilayer Perceptron}
    \label{fig:SiteMLP}
\end{subfigure}
\begin{subfigure}{0.325\textwidth}
    \centering
    \includegraphics[width=\textwidth]{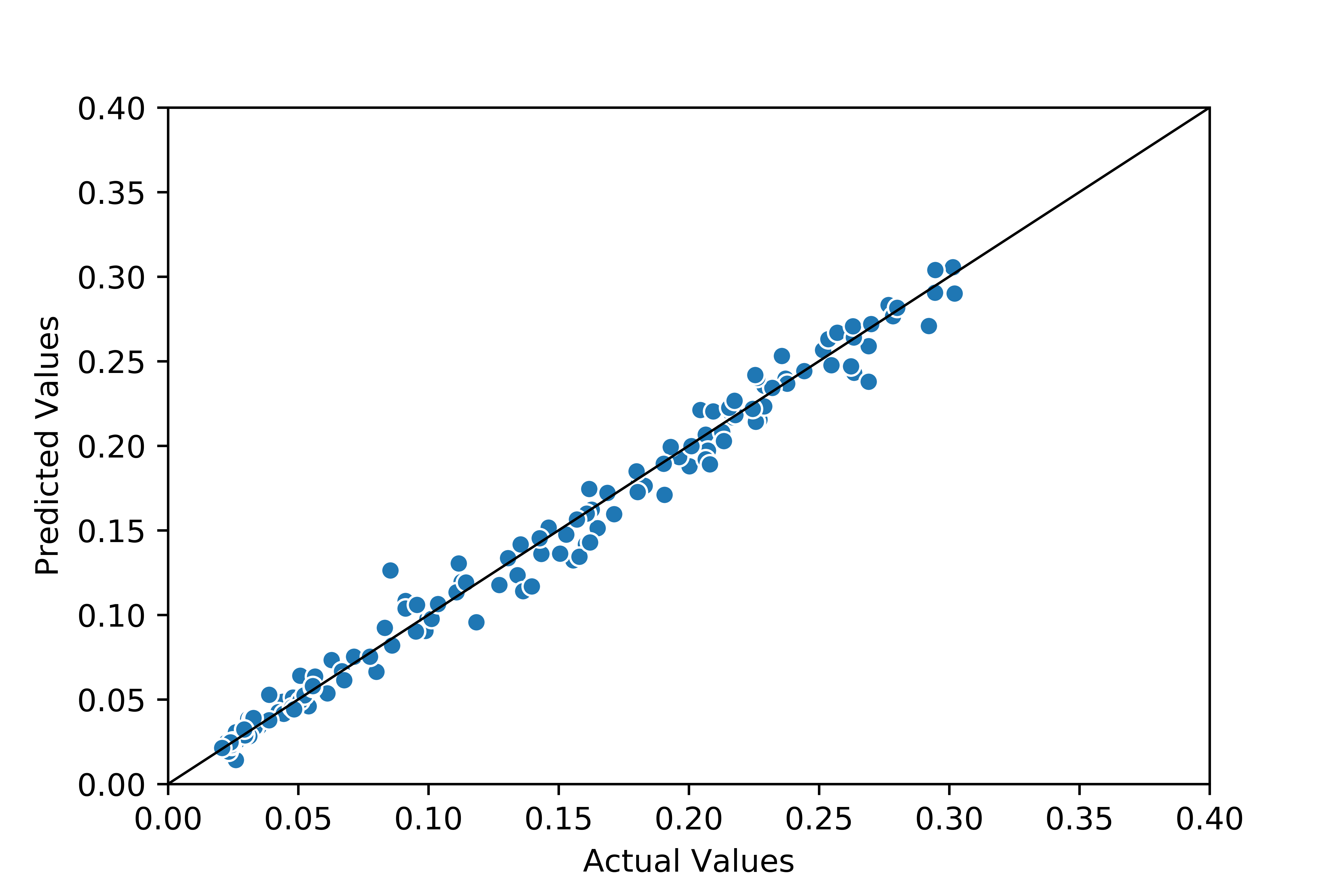}
    \caption{Gradient Boosting Regressor}
    \label{fig:SiteGBR}
\end{subfigure}      
\caption{Performance of the machine learning models in predicting site percolation threshold, as plotted with predicted values in the y-axis and actual values in the x-axis}
\label{fig:Site}
\end{figure*}

% EXPLOSIVE PERCOLATION THRESHOLD RESULTS
\begin{figure*}[t!]
\begin{center}
\begin{subfigure}{0.325\textwidth}
    \includegraphics[width=\textwidth]{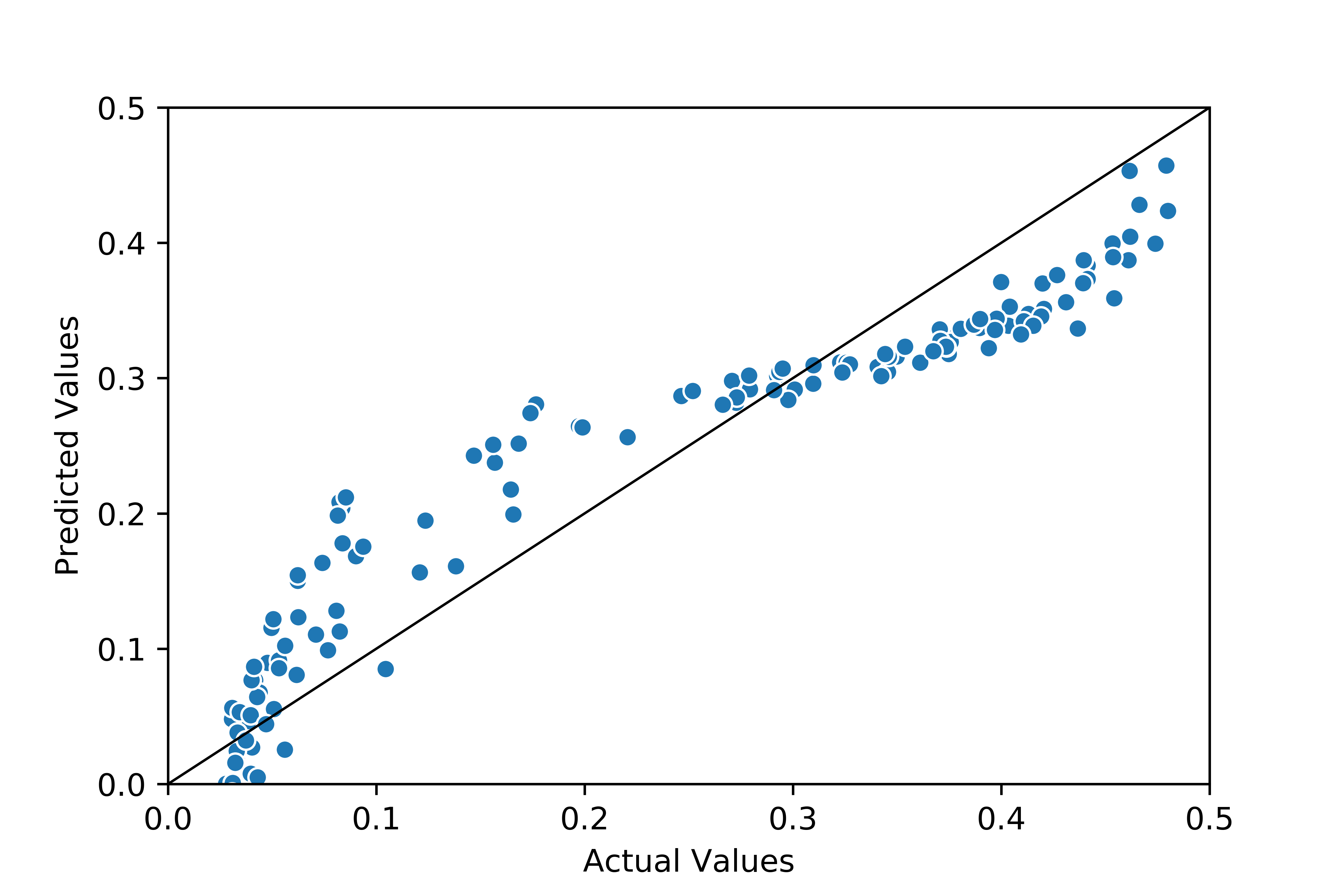}
    \caption{Linear Regression}
    \label{fig:ExplosiveLR}
\end{subfigure}
\hfill
\begin{subfigure}{0.325\textwidth}
    \includegraphics[width=\textwidth]{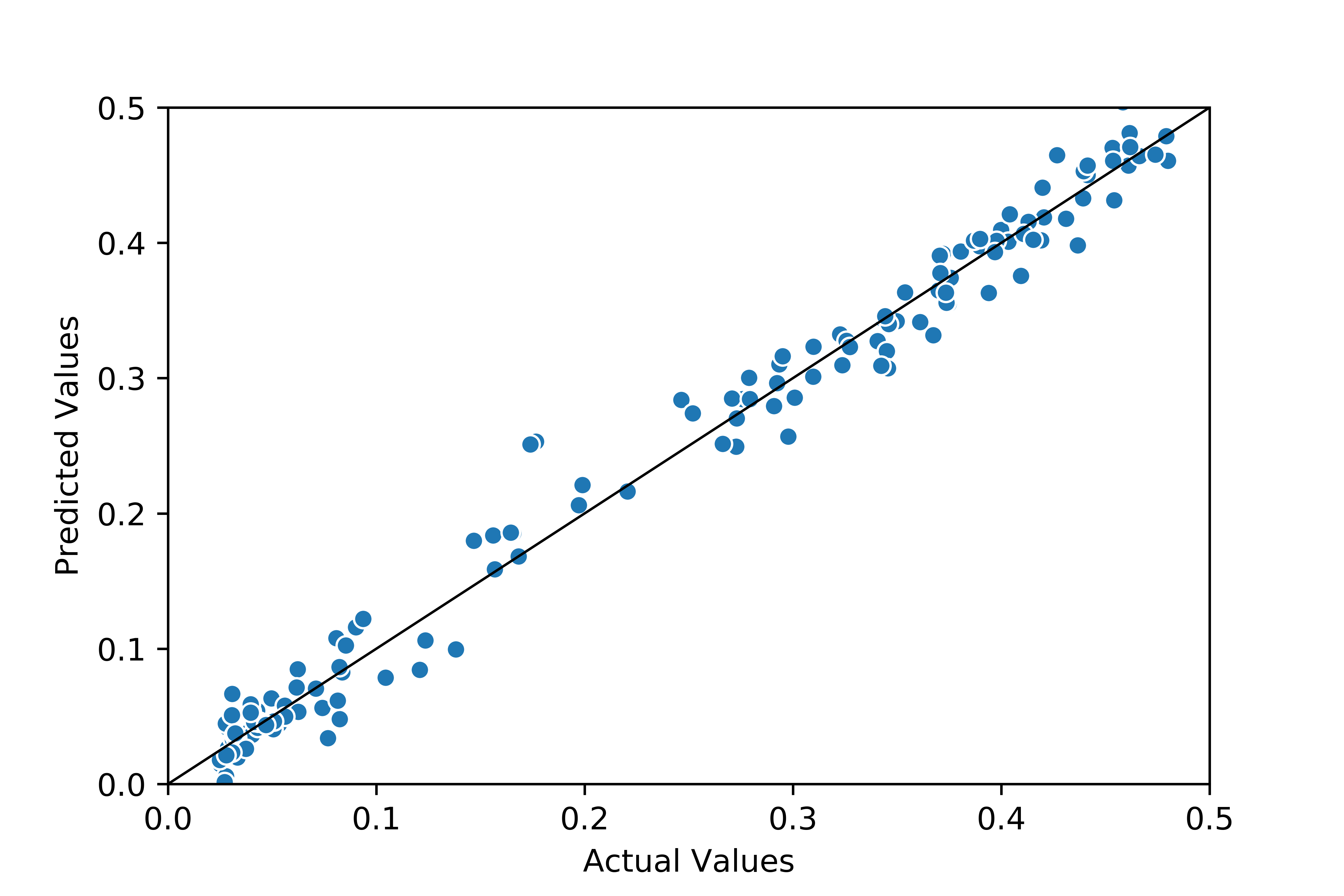}
    \caption{Simple ANN}
    \label{fig:ExplosiveANN}
\end{subfigure}
\hfill
\begin{subfigure}{0.325\textwidth}
    \includegraphics[width=\textwidth]{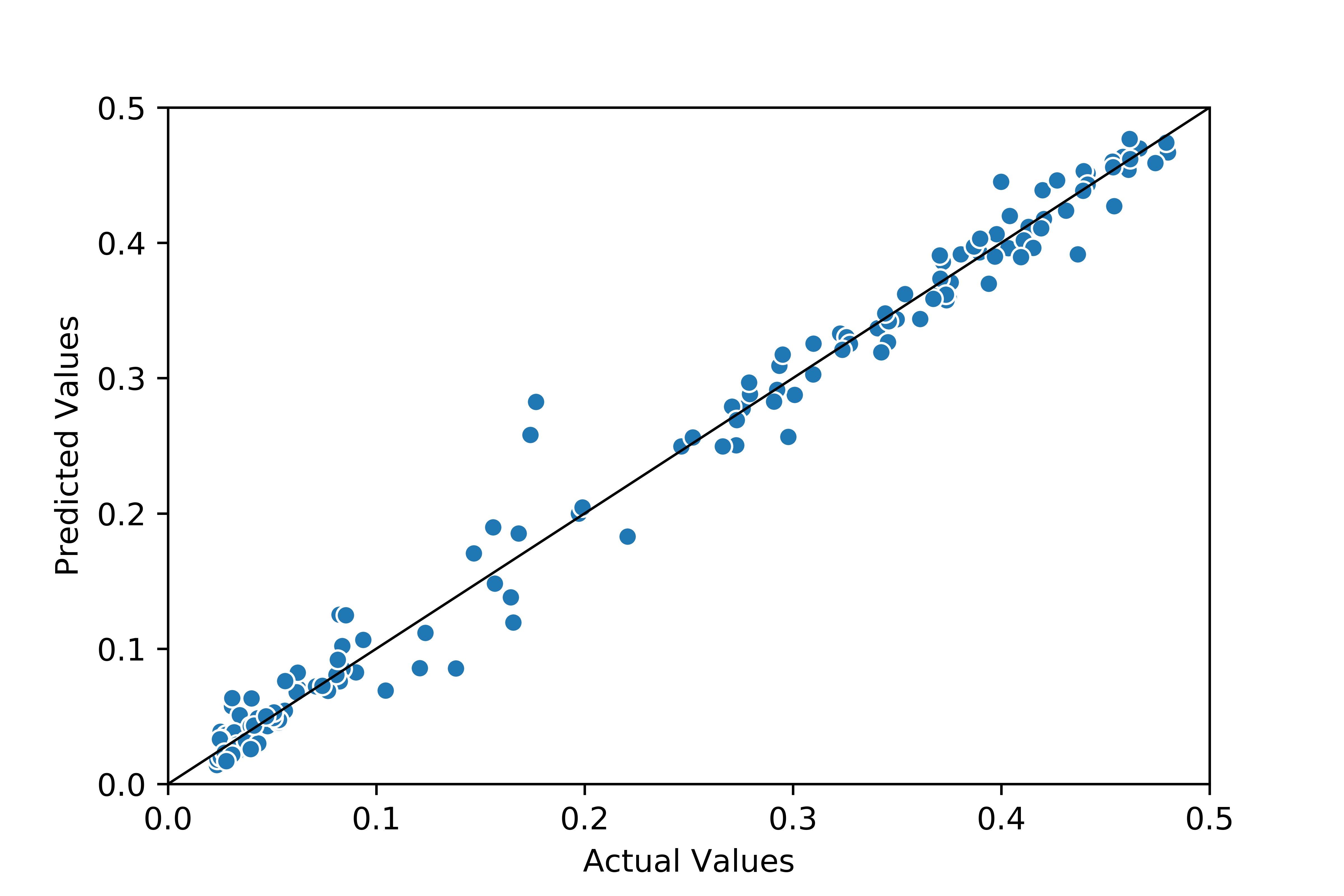}
    \caption{Random Forests Regressors}
    \label{fig:ExplosiveRFR}
\end{subfigure}
\hfill
\begin{subfigure}{0.325\textwidth}
    \includegraphics[width=\textwidth]{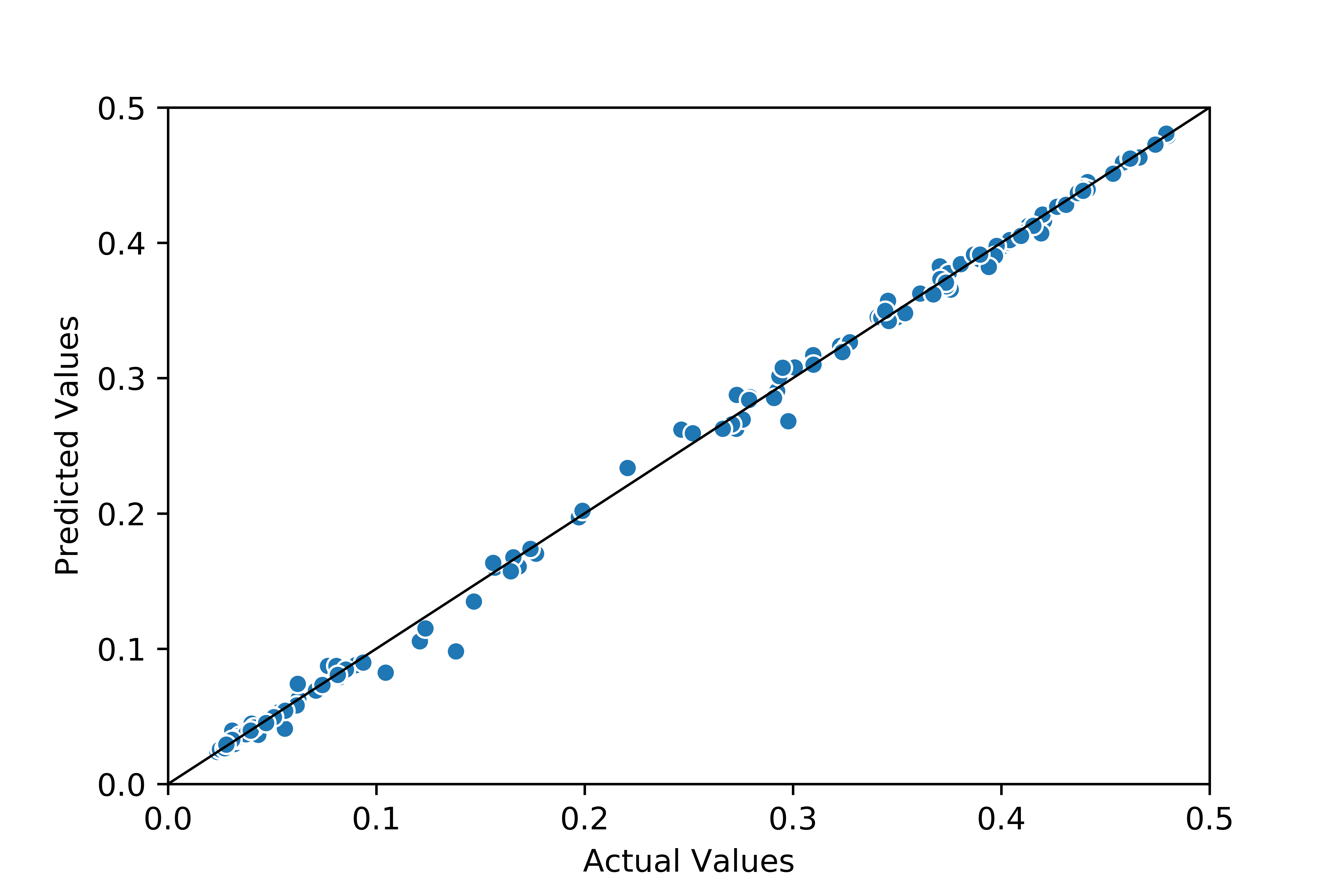}
    \caption{Multilayer Perceptron}
    \label{fig:ExplosiveMLP}
\end{subfigure}
\begin{subfigure}{0.325\textwidth}
    \includegraphics[width=\textwidth]{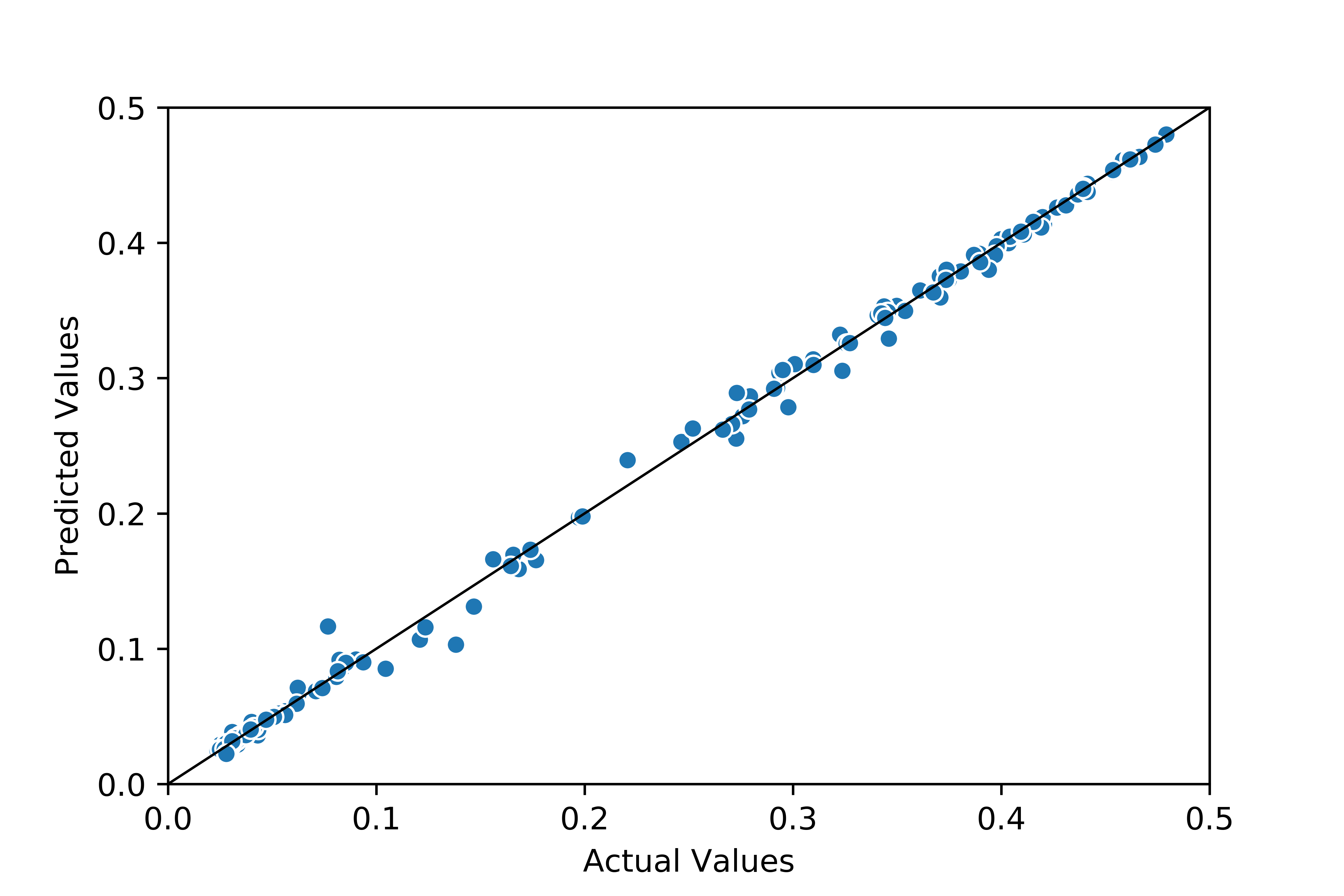}
    \caption{Gradient Boosting Regressor}
    \label{fig:ExplosiveGBR}
\end{subfigure}
\end{center}       
\caption{Performance of the machine learning models in predicting explosive percolation threshold, as plotted with predicted values in the y-axis and actual values in the x-axis}
\label{fig:Explosive}
\end{figure*}
{ For evaluating the accuracy of all models, we have used the root mean squared error as a metric, given as,

\begin{equation}
    RMSE= \sqrt{\frac{1}{T} \sum_{i=1}^{T} {(AV - PV)}^2}
\end{equation}

where, $T$ is the total number of observations, AV and PV signified the actual and predicted values respectively.} Evaluation of the estimators $P_1$, $P_2$ and $P_3$ yields results that have significant error percentages. We find that RMSEs of $P_1$, $P_2$ and $P_3$ for bond percolation threshold are approximately 0.276084, 0.171556 and 0.135102, respectively as enlisted in TABLE \ref{table:tab-empirical}. The Q-Q plots are given in FIG. \ref{fig:estimators}.
The obtained errors in the sparse networks could be partly attributed to the inherent variance in the values of the true percolation threshold. The variance in the computationally obtained values of the percolation threshold over 100 iterations increases as the value of the percolation threshold increases, i.e., in the case of sparse networks, there exists a considerable inherent variance in the point where the phase transition occurs. 
%-----------------------------------------------------------------------
% Feature Importance RESULTS
\begin{figure}
    \includegraphics[width=.99\columnwidth]{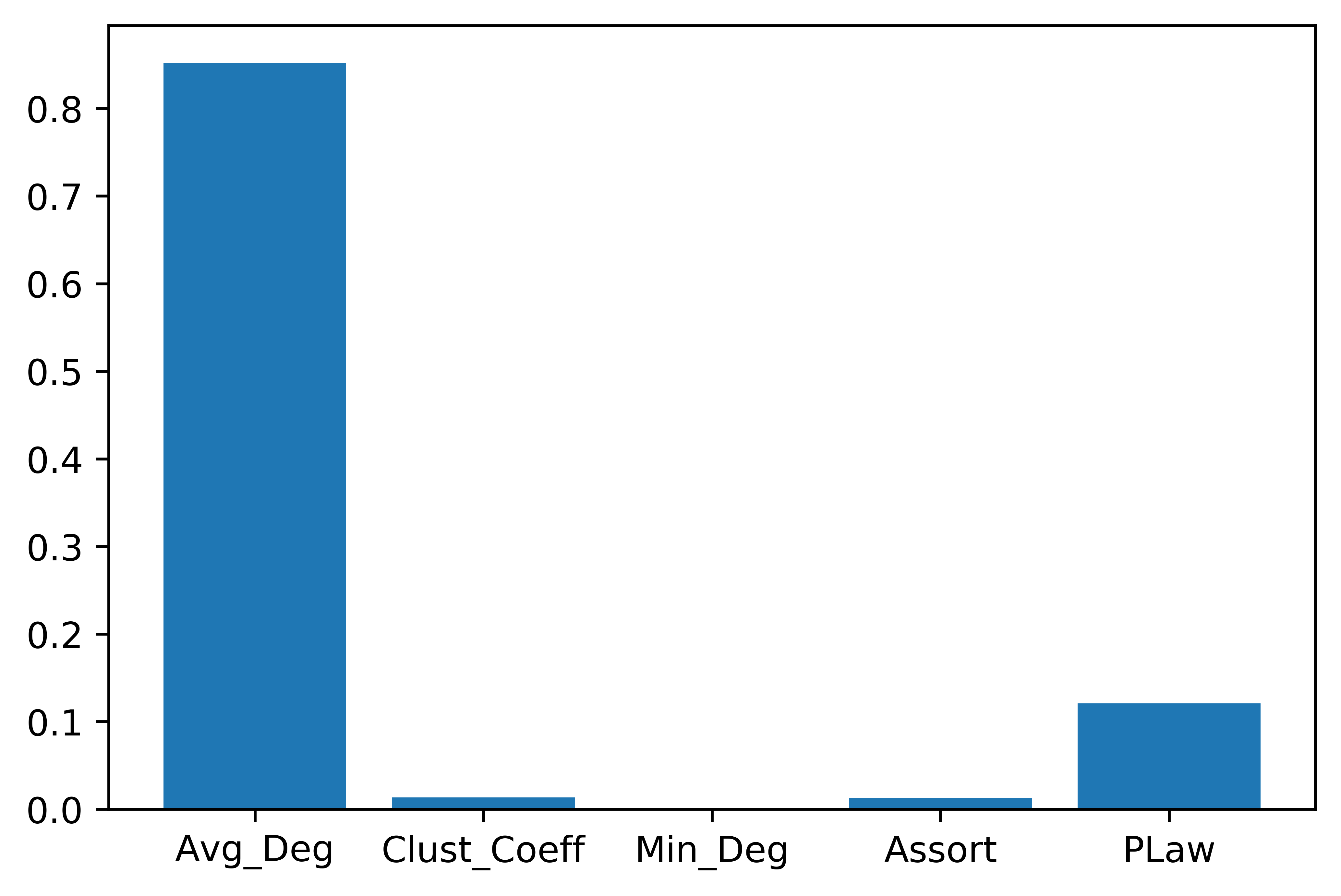}
    \caption{Feature importance investigation of the dataset using Random Forest Regressor method, which shows the average degree of the networks as the most important feature of the used dataset, and minimum degree as the least}
\label{fig:features}
\end{figure}

%----------------------------------------------------------------
%edit: add feature importance selection results
Moreover, as pointed out earlier, estimators $P_2$ and $P_3$ give a lower bound for the value of the percolation threshold. However, the difference between the true $p_c$ and the predicted lower bound increases as networks become more sparse, as is apparent from FIG. \ref{fig:p2} and FIG. \ref{fig:p3}. This may be attributed to the high variance in the values of the true percolation threshold for sparse networks.

It has been observed that some networks undergo phase transition twice \cite{dualpt, mulphase}. In this case, the machine learning models predicts the first transition, as the method used to compute the true value of the percolation threshold considers the size of the second-largest component, which will always be larger for the first phase transition.

The Root Mean Squared Errors for each model observing the percolation threshold for each percolation benchmark is given in TABLE \ref{table:results}.
To train the models, we use $64\%$ of the data as training data, $20\%$ as the test data, and $16\%$ as the validation data. We obtain mean squared errors of $3.8331\times10^{-2}$, $1.5919\times10^{-2}$, $8.652\times10^{-3}$, $8.892\times10^{-3}$ and $7.756\times10^{-3}$ on the test dataset using linear regression, simple ANN, random forests (with important feature selection), MLP and Gradient boosting respectively, in case of the bond percolation threshold. 

In the case of site percolation threshold, we obtain mean squared errors of $3.4795\times10^{-2}$, $1.7044\times10^{-2}$, $8.01\times10^{-3}$, $9.166\times10^{-3}$ and $8.114\times10^{-3}$ on the test dataset using linear regression, simple ANN, random forests (with important feature selection), MLP and Gradient boosting respectively. Finally, we obtain mean squared errors of $6.5667\times10^{-2}$, $4.5706\times10^{-2}$, $8.2\times10^{-3}$, $9.266\times10^{-3}$ and $7.334\times10^{-3}$ on the test dataset using linear regression, simple ANN, random forests, MLP and gradient-boosting respectively, in case of the explosive percolation threshold.

The Q-Q plots for all the predictive models for each case of predicting percolation thresholds are given in FIG. \ref{fig:Bond}, FIG. \ref{fig:Site} and FIG. \ref{fig:Explosive}. The error in prediction for most of the networks in the test data was less than $10^{-2}$ for both the random forests and the ANN-based model. 

Finally, we performed the feature importance to identify the important feature in the dataset using Random Forest Regressor. It is clearly shown in FIG. \ref{fig:features} that the average degree of the networks is the most important feature and the minimum degree is the least important feature of the used dataset.
{ The average degree of the network is found to have a higher degree of impact for predicting the percolation threshold while the minimum degree does not have any effect on the prediction of the percolation threshold.}

\section{Conclusions}
\label{section:conclusion}

% \textit{Justify abstract}
The findings show that machine learning techniques can be used effectively to predict the percolation threshold of networks. The five methods are performed where Gradient Boosting performs best, followed by Random Forest Regressor, Multilayer Perceptron, ANN and Linear Regressor, in estimating percolation on the generated dataset. In particular, the gradient-boosting regressor and random forests models predicted the percolation threshold with very high accuracy. 

We also note that the performance of the machine learning models improves as the quantity and quality of the data improve. Therefore, the machine learning models will perform better as more data becomes available. 

For feature importance, we have used Random Forest Regressors on the dataset for all three types of the percolation threshold. The average degree has a relative importance of ($85\%$) among the features considered, followed by Power-law Degree Distribution ($12\%$) and the least important feature is Minimum Degree. Hence while selecting important features, we drop the Minimum Degree feature. 
%\\explosive percolation \cite{explosive} and 
\\It is also possible to extend this framework to train the machine learning models to predict the percolation threshold of multilayer networks using the same methodology. Furthermore, it is used to predict the percolation threshold of different percolation processes that are important to many other real-world systems, for instance, group percolation \cite{group}. 

Our key argument in the case of bond percolation threshold is that we could display a significant improvement in performance over empirical estimators, with the use of machine learning models. For site and explosive percolation threshold, the usage of machine learning models has a novelty factor. At the implementation stage, we have been able to utilise various machine learning models and neural networks with basic parameters and intermediate levels of network complexity, to assess and compare their results against one another for the specific use case.

\section*{Acknowledgement}
This research received no specific support from governmental, private, or non-profit funding agencies. The content and writing of the paper are solely the responsibility of the authors. Siddharth Patwardhan, Utso Majumder and Aditya Das Sarma are thankful to IISER Kolkata for the project work and access to the DIRAC supercomputing facility. The authors Mayukha Pal and Divyanshi Dwivedi wish to thank ABB Ability Innovation Centre, India for their support in this work.

\section*{CRediT authorship contribution statement}

\textbf{Siddharth Patwardhan:} Conceptualization, Methodology, Software, Data curation, Writing- Original draft. \textbf{Utso Majumder:} Methodology, Software, Data curation, Writing- Original draft. \textbf{Aditya Das Sharma:} Software, Data curation. \textbf{Mayukha Pal:} Methodology, Validation, Supervision, Writing- Reviewing and Editing. \textbf{Divyanshi Dwivedi:} Writing- Reviewing and Editing. 
\textbf{Prasanta K. Panigrahi:} Conceptualization, Methodology, Project administration, Validation, Supervision, Writing- Reviewing and Editing.

\section*{Declaration of Competing Interest}

The authors declare that they have no known competing financial interests or personal relationships that could have appeared to influence the work reported in this paper.

\bibliographystyle{apsrev4-1}  

\bibliography{main}

\end{document}